\documentclass[a4paper,fleqn,usenatbib]{mnras}

\usepackage{graphicx}	
\usepackage{epstopdf}
\usepackage{ulem}
\usepackage{xcolor}

\title[NIR metallicity map of the SMC]{The VMC survey -- XXXIX: Mapping metallicity trends in the Small Magellanic Cloud using near-infrared passbands}

\author[S. Choudhury et al.]{Samyaday Choudhury$^{1,2}$\thanks{E-mail: samyaday.choudhury@gmail.com},
Richard de Grijs$^{1,2,3}$,
Stefano Rubele$^{4,5}$,
Kenji Bekki$^{6}$,
\newauthor {Maria-Rosa L. Cioni$^{7}$, 
Valentin D. Ivanov$^{8}$,
Jacco Th. van Loon$^{9}$},
\newauthor {Florian Niederhofer$^{7}$,
Joana M. Oliveira$^{9}$,
and Vincenzo Ripepi$^{10}$}\\
$^{1}$ Department of Physics and Astronomy, Macquarie University, Balaclava Road, Sydney, NSW 2109, Australia\\
$^{2}$ Research Centre for Astronomy, Astrophysics and Astrophotonics, Macquarie University, Balaclava Road, Sydney, NSW 2109, Australia\\
$^{3}$ International Space Science Institute--Beijing, 1 Nanertiao, Zhongguancun, Hai Dian District, Beijing 100190, China\\
$^{4}$ Dipartimento di Fisica e Astronomia, Universit\`a di Padova, Vicolo dell'Osservatorio 2, I-35122 Padova, Italy\\
$^{5}$ Osservatorio Astronomico di Padova, INAF, Vicolo dell'Osservatorio 5, I-35122 Padova, Italy\\
$^{6}$ ICRAR, M468, University of Western Australia, 35 Stirling Hwy, 6009 Crawley, Western Australia, Australia\\
$^{7}$ Leibniz-Institut f\"{u}r Astrophysik Potsdam (AIP), An der Sternwarte 16, D-14482 Potsdam, Germany\\
$^{8}$ European Southern Observatory, Karl-Schwarzschild-Str. 2, D-85748 Garching bei M\"{u}nchen, Germany\\
$^{9}$ Lennard-Jones Laboratories, Keele University, ST5 5BG, UK\\
$^{10}$ INAF--Osservatorio Astronomico di Capodimonte, via Moiariello 16, I-80131, Naples, Italy\\
} 
\date{Accepted XXX. Received YYY; in original form ZZZ}

\pubyear{2015}

\begin{document}
\label{firstpage}
\pagerange{\pageref{firstpage}--\pageref{lastpage}}
\maketitle

\begin{abstract}

We have derived high spatial resolution metallicity maps covering $\sim$42 deg$^2$ across the Small Magellanic Cloud (SMC) in an attempt to understand its metallicity distribution and gradients up to a radius of $\sim$ 4$^{\circ}$. Using the near-infrared VISTA Survey of the Magellanic Clouds, our data cover a thrice larger area compared with previous studies. We identify red giant branch (RGB) stars in spatially distinct $Y, (Y-K_{\rm s})$ colour--magnitude diagrams. In any of our selected subregions, the RGB slope is used as an indicator of the average metallicity, based on calibration to metallicity using spectroscopic data. The metallicity distribution across the SMC is unimodal and can be fitted by a Gaussian distribution with a peak at [Fe/H] = $-$0.97 dex ($\sigma$[Fe/H] = 0.05 dex). We find evidence of a shallow gradient in metallicity ($-0.031 \pm 0.005$ dex deg$^{-1}$) from the galactic centre to radii of 2$^{\circ}$--2.5$^{\circ}$, followed by a flat metallicity trend from $\sim$ 3.5$^{\circ}$ to 4$^{\circ}$. We find that the SMC’s metallicity gradient is radially asymmetric. It is flatter towards the East than to the West, hinting at mixing and/or distortion of the spatial metallicity distribution (within the inner 3$^{\circ}$), presumably caused by tidal interactions between the Magellanic Clouds.

\end{abstract}



\section{Introduction}
The spatial distribution of heavy elements within a galaxy traces its mass accumulation history (\citealt{Ho+2015}; and references within), whereas the presence of a metallicity gradient (MG) within a galaxy provides important indications regarding its evolution and interaction history. The Magellanic Clouds (MCs) represent a system that is composed of two interacting galaxies, the Large and Small Magellanic Clouds (LMC, SMC), located at distances of about 50 and 60 kpc, respectively. The MCs, along with the Milky Way (MW), form the closest example of an interacting system of galaxies (\citealt{Murai&Fujimoto1980theMagStream, Tanaka1981theMagStream, Fujimoto&Murai1984theMagStream, Gardiner+1994numsimul, Westerlund1997theMCs}), thus making the estimation of their mean metallicities and any radial MGs important to understand the system's evolution history. 

Results from {\sl Hubble Space Telescope} proper motions \citep{Kallivayalil+2006a,Kallivayalil+2006b,Kallivayalil+2013} suggest that the LMC and SMC are on their first infall trajectory towards the MW, or perhaps they are orbiting the Galaxy with a long period \citep[$>$6 Gyr;][]{Besla+2010ApJsimulationsofMS}. Simulations suggest that the origin of various dynamical features seen in the Magellanic System, e.g., the Magellanic Bridge or Stream, may be caused by periodic interactions between the LMC and SMC \citep{Diaz&Bekki2011MNRASconstrain,Besla+2012MNRAStherole}.  
The scenario for the SMC is interesting, since its outskirts are dominated by its dynamical interaction history with the LMC and/or the MW. 
The SMC is gas rich, with poorer metallicity ($Z \approx 0.004$) compared with the LMC ($Z \approx 0.008$) and the MW ($Z \approx 0.02$), and it hosts active star-forming regions like the very prominent N\,66/NGC\,346 \citep{Cignoni+2011AJ} and thousands of young stellar objects \citep{Sewilo+2013ApJ}. The mutual interaction between the MCs, rather than their interaction with the MW, can be fundamental in shaping their star-formation history and any MGs \citep{Cioni2009A&Athemetallicity}. Signatures of tidal distortions are observed in the SMC's outer regions ($>$ 2--2.5 deg), in gas as well as in its stellar populations \citep{Nidever+2013,Smitha+2017MNRAS}. Periodic interactions with the LMC \citep{Diaz&Bekki2011MNRASconstrain,Besla+2012MNRAStherole} can also affect the mean metallicity or radial MG within the galaxy. Thus, accurate measurement of the MGs over a large radial extent from the galaxy's centre can provide insights to understand galaxy formation and evolution mechanisms in an interacting environment. 

There has been little consensus between spectroscopic and photometric estimates about the possible presence of a radial MG across the SMC over the past decade. Previous spectroscopic estimates using a few hundred stars (within small pockets in the SMC) to a few thousand stars (spanning relatively wider areas) report a significant MG (see values below).  Most of these studies used Calcium II triplet (CaT) spectroscopic metallicity measurements, an excellent tool to estimate metallicities of Red Giant Branch (RGB) stars in the MCs \citep[e.g.,][]{Cole+2005AJspectroOfRGs, Grocholski+2006AJCaIItriplet, Carrera+2008AJ-CEH-SMC, Dobbie+2014MNRAS-papI, Parisi+2016AJfieldII}.  \cite{Carrera+2008AJ-CEH-SMC} used CaT spectroscopy of more than 350 RGBs distributed in 13 fields located at different positions across the SMC (ranging from 1$^{\circ}$ to 4$^{\circ}$ from the galaxy's centre). They estimated a mean [Fe/H] $\sim -1.0$ dex within the inner SMC and a decreasing mean metallicity for the two outermost fields. \cite{Dobbie+2014MNRAS-papI,Dobbie+2014MNRAS-papII} carried out the most extensive spectroscopic study of RGB stars in the SMC. They studied about 3000 giants within the SMC's inner 5$^{\circ}$. \cite{Dobbie+2014MNRAS-papII} [henceforth D14] confirmed a median metallicity, [Fe/H] $= -0.99 \pm 0.01$ dex and a MG of $-0.075 \pm 0.011$ dex deg$^{-1}$. More recently \cite{Parisi+2016AJfieldII} increased the sample of RGB stars ($\sim$750) from that studied in their previous study \citep{Parisi+2010AJfieldI} and estimated a median metallicity of [Fe/H] $= -0.97 \pm 0.01$. Their results agree with that of D14 in the sense that they detected a MG of $-0.08 \pm 0.02$ dex deg$^{-1}$ within the inner 4$^{\circ}$. Very recently, \cite{Nidever+2020ApJ} reported metallicities of 3600 RGBs in the MCs based on high-resolution $H$-band spectra from the Apache Point Observatory Galactic Evolution Experiment (APOGEE) survey. Their work aimed at estimating the $\alpha$-element abundances and the star-formation efficiency within the MCs and not at determining radial MGs. 
 
Photometric data can be used to understand global variations in metallicity out to greater distances from the galactic centre, thus covering larger area as compared to spectroscopic data. However, most previous photometric studies that derived the SMC's metallicity distribution reported a constant metallicity with radius or, at best, a relatively shallow MG in contrast to the above mentioned spectroscopic estimates. Estimation of the MGs also depends on the age and metallicity of the population used as a tracer. \cite{Cioni2009A&Athemetallicity} used the ratio of carbon-rich (C-type) to oxygen-rich (M-type) field asymptotic giant branch (AGB) stars, which is a proxy of [Fe/H], to estimate an almost constant ($-1.25 \pm 0.01$ dex) metallicity distribution to a galactocentric distance of 12 kpc ($\approx$ 11.5$^\circ$). However, their indicators (AGB) and calibrators (RGB stars) were different. \cite{Piatti2012MNRASage-metalSMC} analysed Washington ($CT_{1}$) photometric data of about 3.3 million field stars distributed throughout the entire SMC main body to investigate the field's age--metallicity relationship from the formation of the galaxy until $\sim$1 Gyr ago. Their investigation led them to conclude that the field stars do not exhibit any gradients in age and/or metallicity. Recent studies of old variable stars based on large-area photometric survey data of the SMC do not report any MG \citep{Kapakos+2011MNRAS, Haschke+2012aAJ, Deb+2015MNRAS} or at most a small tendency of increasing metal abundance towards the SMC's centre \citep{Kapakos&Hatz2012MNRAS}.

In an attempt to understand the metallicity variation within the SMC's inner radial range, $< 2.5^{\circ}$, \citet[][hereinafter C18]{C18} created the first high spatial resolution metallicity map based on RGB stars, by combining large area photometric data (Magellanic Cloud Photometric Survey \citep[MCPS;][]{Zaritsky+2002AJMCPSSMC} and \citep[OGLE III;][]{Udalski+2008AcAOGLEIIISMC} with spectroscopic data of RGB stars. The mean SMC metallicity estimated by C18 is [Fe/H] $= -0.94$ dex ($\sigma$[Fe/H] = 0.09 dex) for OGLE III, and [Fe/H] $= -0.95$ dex ($\sigma$[Fe/H] = 0.08 dex) for the MCPS. They confirmed a gradual MG to a radius of $\approx$ $2.5^{\circ}$ ($-0.045 \pm 0.004$ dex deg$^{-1}$ for MCPS and $-0.067 \pm 0.006$ dex deg$^{-1}$ for OGLE III), a little shallow with respect to spectroscopic studies. However, optical passbands are susceptible to the effects of differential reddening, particularly in the inner regions of the galaxy. C18 could not estimate metallicity trends in areas around the SMC centre ($<$1.0$^{\circ}$) or in the North East, possibly because of variable reddening and/or line-of-sight (LOS) depth variations. To properly isolate a global MG, it is important to understand metallicity variations across those areas, out to radii $>$2.0$^{\circ}$--2.5$^{\circ}$, where the interaction signatures with the LMC become significant \citep{Nidever+2013,Smitha+2017MNRAS}. 

C18 used the technique employed by \citet[][henceforth C16]{Choudhury+2016MNRAS} for a similar study of the LMC. The RGB was identified in the $V$ versus $(V - I)$ colour--magnitude diagrams (CMDs) of small regions of varying sizes within the SMC. The estimated slope of the RGB was used as an indicator of the mean metallicity of a small region. The relationship between RGB slope and metallicity is well-known and was shown by \cite{DaCosta&Armandroff1990AJstandard} for the first time using six Galactic globular clusters observed in the Cousins $V$ and $I$ bands. Since then, the dependence of the RGB slope and the RGB's morphology on metallicity has been demonstrated by several other studies using near-infrared (NIR) passbands \citep{Frogel+1983,Kuchinski+1995AJ-IRarray, Tiede+1997, Ivanov+2000, Ivanov&Borissova2002}, since RGB stars are brighter at NIR wavelengths than in the optical. NIR passbands are less sensitive to metallicity compared with optical filters. However, the RGB resembles a straight line much more closely in NIR than in optical CMDs, thus making slope estimations easier. The above studies were confined to globular and open clusters, which are internally homogeneous in stellar ages and metallicities. However, the field stellar population is heterogeneous. Its RGB will not consist of a single population, and the dominant population dictates the shape as well as the slope of the RGB (C16, C18). Thus, the RGB slope of a field region will correspond to the metallicity of the area's dominant RGB population. The technique of C16 identifies the dominant population and estimates its slope for field populations. The RGB slopes are then calibrated to metallicity using spectroscopic data for field RGB stars in selected small regions.

This paper goes a step further towards understanding the spatial metallicity distribution. We explore whether a MG is present based on high-resolution spatial bins out to a radial distance of $\sim$ 4$^\circ$ from the optical centre of the SMC, using the NIR passbands from the VISTA Survey of the Magellanic Clouds \citep[VMC;][]{Cioni+2011}. The VMC survey covers $\sim$ 2--3 times larger area compared with the OGLE III and MCPS surveys. Moreover, NIR passbands are relatively less affected by differential reddening compared with optical bands. We use the technique developed by C16 (and used by C18), modified for the NIR passbands of the VMC to derive a metallicity map of the SMC. This paper is organised as follows. In Section 2 we describe the VMC data. Our analysis of the RGB slope estimation and calibration to metallicity are presented in Section 3. Section 4 contains our main results, the NIR metallicity maps of the SMC. A discussion of the results is presented in Section 5. We summarise our conclusions in Section 6.

\section{Data} 

The VMC survey is a uniform and homogeneous survey of the Magellanic System in NIR passpands using the 4 m VISTA telescope at La Silla Paranal Observatory, Chile, and is one of the European Southern Observatory's (ESO) public surveys. The telescope is equipped with the VISTA infrared camera (VIRCAM), an array of 16 Raytheon detectors with a mean pixel size of 0.339 arcsec and 1.65 degree diameter field of view \citep{Sutherland+2015}. Observations began in 2009 and were completed by 2018, covering about $\sim$170 deg$^2$ of the Magellanic System (LMC: $\sim$ 105 deg$^2$; SMC: $\sim$ 42 deg$^2$; Magellanic Bridge: $\sim$ 20 deg$^2$; Magellanic Stream: $\sim$ 3 deg$^2$). The survey uses three NIR passbands, $Y, J$ and $K_{\rm s}$, centred on $\lambda = 1.02, 1.25$ and $2.15 \mu$m, and reaches a 5$\sigma$ magnitude limit of $Y = 21.9, J = 22$, and $K_{\rm s} = 21.5$ mag in the Vega system. A single tile represents a mosaic of six paw-print images in a given passband ($YJK_{s}$). The number of such tiles covering the LMC, SMC, Magellanic Bridge and Stream are 68, 27, 13 and 2, respectively. We direct readers to \cite{Cioni+2011} for a detailed description of the survey and its science goals.

Each of the 27 tiles covering the SMC cover almost uniformly an area of 1.5 deg$^2$ \citep[by a minimum of two pixels;][]{Sutherland+2015} and their centres extend out to 3.5$^\circ$--4$^\circ$ from the SMC’s centre. In this study, we use the point spread function (PSF) photometry catalogue of \cite{Rubele+2018MNRAS}. They retrieved VMC data from the VISTA Science Archive \citep[VSA;][]{Cross+2012A&A}, and performed PSF analysis using already processed and calibrated paw-print images provided by the VISTA Data Flow System \citep[VDFS;][]{Emerson+2004SPIE, Irwin+2004SPIE}. They homogenised individual paw-print PSFs, and then combined them into deep tile images on which the PSF photometry was performed; see their Section 2 and \cite {Rubele+2015} for details on the methodology.

\citet[][their appendix]{Rubele+2015} include a figure showing the objects' sharpness criteria as a function of magnitude for the VMC passbands. Sources with sharpness $< -1$ may be bad pixels, whereas objects characterised by a sharpness parameter $> +1$ could be extended sources (which we avoid here). We
also only consider stars with photometric uncertainties $\le 0.15$ mag. \cite{Rubele+2018MNRAS} obtained PSF photometry down to the 50 per cent completeness limits in all tiles, corresponding to magnitude limits of $Y=21.25, J=20.95$, and $K_{\rm  s}=20.45$ mag. We use Red Clump (RC) stars and the upper section of the RGB in this study, which corresponds to a magnitude range that is about 2--6 mag brighter than the $Y$-band 50 per cent completeness limit. Overall the $Y$-band photometry has a typical uncertainty of 0.05 mag, for stars brighter than $Y=$ 21 mag.


\section{Analysis}

\subsection{Estimation of the RGB slopes}
\begin{figure*} 
\includegraphics[height=4.0in,width=5.0in]{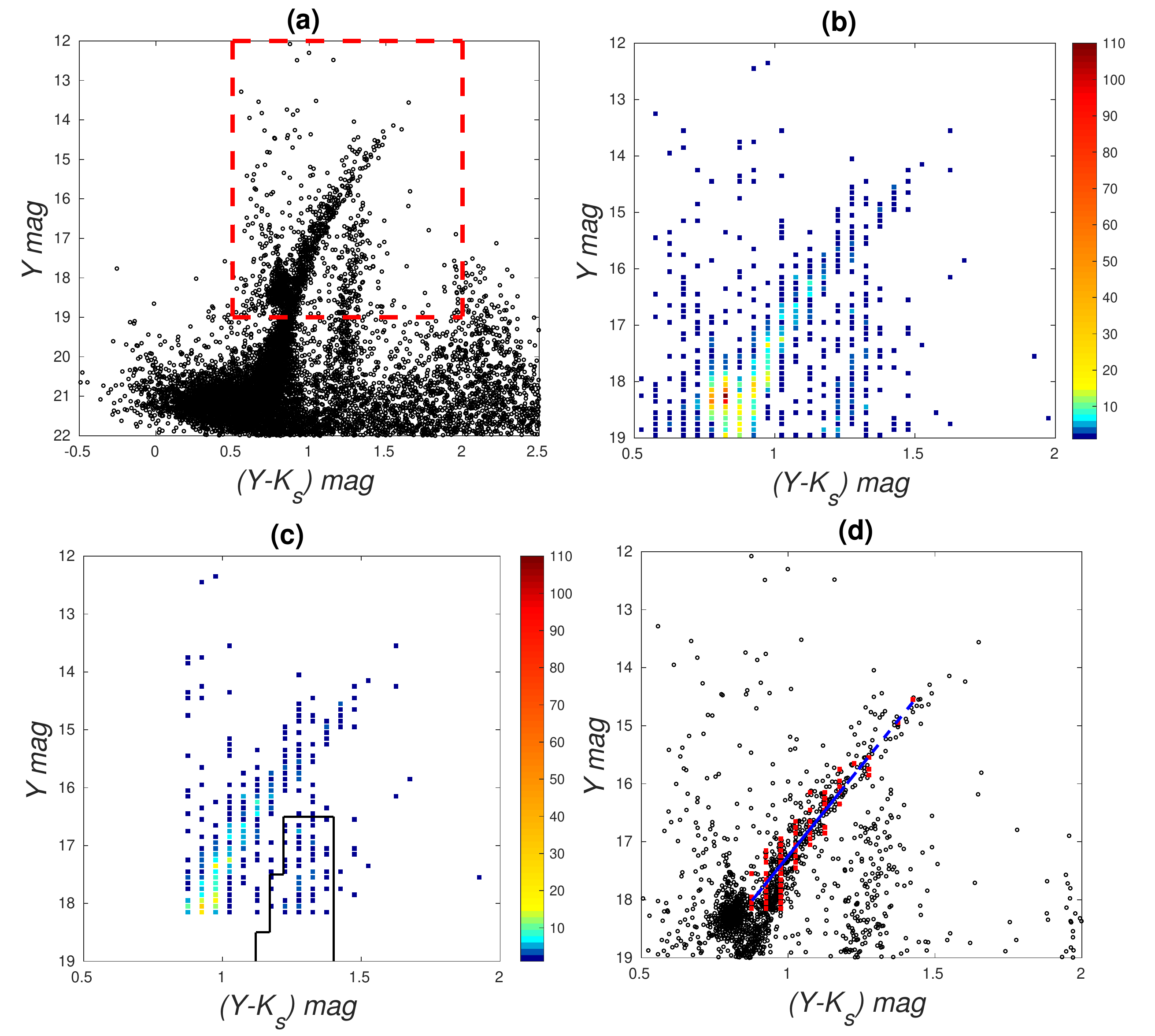}
\caption {(a) $Y$ versus $(Y-K_{\rm s})$ CMD of a 15.70$\times$16.20 arcmin$^2$ SMC subregion at RA = $9.95^\circ$, Dec = $-71.77^\circ$, containing $N=7499$ stars (black open circles). Stars in the rectangle (red dashed line) belong to the evolved part of the CMD. (b) Density diagram of the evolved part of CMD, where the bins are colour-coded based on stellar numbers (see the colour bar). (c) Density diagram following application of a colour--magnitude cut at the peak of the RC distribution. The stepped block (black solid lines) was defined to remove MW contamination. (d) CMD of the subregion (black open circles) overplotted with bins containing $\ge$3 stars each (red filled squares). A linear fit to these bins representing the RGB is shown as the blue  dashed line. The colour bars in (b) and (c) represent number of stars in each colour-magnitude bin. The estimated parameters are: $|$slope$| = 6.24 \pm 0.33$, $r =0.93$, and $N_p = 59$.}
\label{fig:lsqfit}
\end{figure*}
\begin{figure*} 
\begin{center} 
\includegraphics[height=4.0in,width=5.0in]{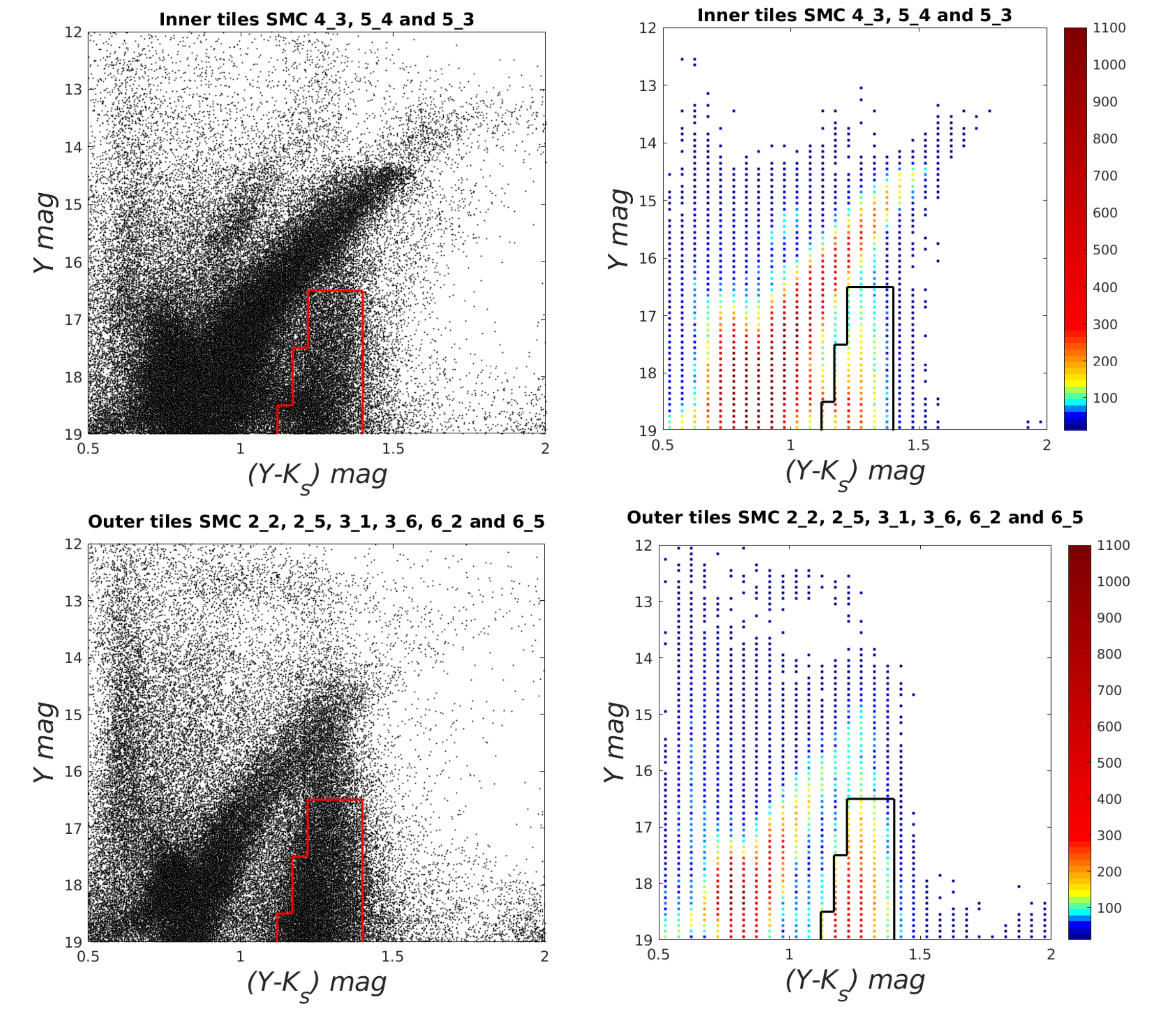}
\caption {Region selected to remove the MW's contamination (stepped block, see Figure \ref{fig:lsqfit}(c)). Top right and top left panels show combined CMD and corresponding number density for three central tiles, respectively. Bottom left and bottom right panels show combined CMD and corresopnding number density for six non-central regions (different directions from the centre), respectively. The colour bars have the same description as Figure \ref{fig:lsqfit}.}
\label{fig:mwcut}
\end{center} 
\end{figure*}
We selected the two extreme VMC passbands, i.e., the $Y$ and $K_{\rm s}$ bands, thus affording us access to the longest colour baseline. This ensures that we have access to the maximum observable effect of metallicity on the upper RGB (the RGB segment brighter than the RC), and hence on the RGB slope. This was checked by overplotting $PARSEC$ isochrones \citep{Bressan+2012MNRAS} on CMDs using different combinations of VMC filters, for a range of metallicities ($Z = 0.004\pm0.002$). The SMC is spatially binned into small regions (subregions, explained later on). The $Y$ versus $Y-K_{\rm s}$ CMDs of these subregions clearly show the presence of main-sequence (MS), RGB and RC stars, as well as other stellar evolutionary stages (Figure \ref{fig:lsqfit}a) .

We adopted the technique used by C16 and C18 for the LMC and SMC, respectively, to estimate the RGB slope in the CMDs of subregion across the SMC. A salient feature of the technique is that we adopt the densest part of the RC in the relevant CMD as the base of the RGB. Since the CMD loci of RC and RGB stars are similarly affected by reddening, the technique can be automated for the entire SMC, irrespective of the reddening. For a detailed description of the technique, readers are directed to Section 3 of C16. The main difference of our current approach with respect to that adopted by C16
is found in the removal of MW contamination. \citet[][their figure 2]{Rubele+2018MNRAS} showed that, apart from SMC main-body features, these CMDs are also affected by MW contamination in $K_{\rm s}$ versus $Y-K_{\rm s}$. A similar effect is noted even in the $Y$ versus $Y-K_{\rm s}$ CMDs. We briefly describe our approach here.

\begin{enumerate}
\item The number of stars in any of our subregion is denoted $N$. We exclude the MS and consider mainly the evolved portion of the CMD, $0.5 < (Y-K_{\rm s}) \le 2.0$ mag and $12.0 \le Y < 19.0$ mag (see Figure \ref{fig:lsqfit}a). 
\item We construct a density distribution to identify the loci of RC stars and adopt the peak in the RC's $(Y-K_{\rm s})$ colour and $Y$
  magnitude as the base of the RGB (see Figure \ref{fig:lsqfit}b). 
\item The bluer and fainter bins (with respect to the RC peak) are removed by a cut in colour and magnitude corresponding to the RC peak (see Figure \ref{fig:lsqfit}c). The CMD is dominated by RGB stars, although it is also contaminated to some extent by some other evolutionary phases (AGB stars, red supergiants, etc.) and by the foreground MW. The MW contamination has two features \citep[see figure 2 of][]{Sun+2018ApJ}. The feature which is bluer than the RGB and resembles a vertical bright strip near the vicinity of the RC distribution is removed by the colour--magnitude cut mentioned in previous point. The other feature is observed as a vertical strip in the redder part of the CMD. To check how the MW contamination affects the CMDs in the inner and outer regions of the SMC, we constructed CMDs by combining a few inner tiles (tiles SMC 4$\_$3, 5\_3 and 5\_4), and a few outer tiles in different directions from the SMC's centre (tiles SMC 2$\_$2, 2\_5, 3\_1, 3\_5, 6\_2 and 6\_5; see Figure \ref{fig:mwcut}, top left and bottom left). The redder MW contamination has a similar density as the  densest portion of the RGB in the density distributions of both CMDs (see Figure \ref{fig:mwcut}, top right and bottom right). To decontaminate this feature, we selected points in the MW-contaminated area of the CMDs which appeared to be of similar density as the densest part of the RGB (see the stepped block). The overall region was chosen carefully, so as to avoid removing portions of the RGB close to the base or the tip and, more importantly, so that the selection can be automated consistently across the entire SMC body. This block is similar to Region $F$ in the $(K_{s}, J-K_{s})$ CMD of \cite{Dalal+2019MNRAS}. According to these authors, about 90 per cent of stars in this region originate from the MW. Thus, once we have removed the bins within this block, the dominant contribution in the density CMD is from the RGB region.
\item Finally, to identify the RGB unambiguously and reduce the scatter in the RGB, eliminate other evolutionary phases and any remaining MW contamination, we considered only those colour--magnitude bins that contained at least three stars ($N_p$). This criterion eliminates the brighter part of the RGB, and typically leads to sampling of the RGB from the RC peak up to 3--3.5 mag brighter in most subregions. We also tried five stars as the minimum number, but this limits the extent of the RGB considered, at the bright part, which in general is poorly populated (C16). Adopting a fewer than three stars can cause contamination from other evolutionary phases e.g. AGB stars, as well as from the MW. Table B2 of \cite{Dalal+2019MNRAS} shows that $\sim$ 14.8 per cent of stars in the upper RGB region are from the MW, implying that the RGB is still the dominant population there. Since we do not perform a weighted fit, the individual content of the bins will not affect our estimated RGB slope. MW decontamination may lead to removal of $\sim$ 1--2 bins near the RGB tip, as those bins are less densely populated compared with bins near the RGB base. This may result in sampling a reduced RGB length (by 0.25--0.5 mag). However, any resulting variation in RGB slope is well within the 1$\sigma$ uncertainties. Also, as we will show below, the RGB slope range and distribution remain unaffected by such length variations (Section 3.2, Figure \ref{fig:para_rc1rc2}). In this study we are interested in the relative variation in RGB slope (and, hence, in metallicity) among subregions rather than accurate estimations for individual subregions.
Figure \ref{fig:lsqfit}(d) shows that the selected part of the RGB appears to be more or less linear, except for the curved, brighter part (the RGB tip region). These bins, which represent the RGB, are fitted with a straight line and the slope ($|$slope$|$$\pm$$\sigma_{\rm slope}$) is estimated using least-squares minimisation (using 3$\sigma$ clipping in a single iteration). The correlation coefficient of the fit (also for other fits in the rest of the paper) is assessed by its absolute value, $r$.
\end {enumerate}

Following C18, we spatially subdivide the SMC tiles further, based on stellar density and not into equal areas. This is because very large $N_p$ ($> 100$) or small $N_p$ ($< 20$) leads to poor value estimations of $|$slope$|$ with $r < 0.5$. Subregions exhibiting large $N_p$ (higher stellar densities) are generally found in the central regions (e.g., tiles SMC 4\_3, 5\_3 and 5\_4), whereas the subregions with small $N_p$ (low stellar densities) are generally found in the outer tiles (e.g., tiles SMC 2$\_$5, 3\_6, 3$\_$1, 4$\_$1, 4\_6, 5\_6, 6\_5, 7\_3 and 7\_4). Inspection of subregions with high stellar density suggests that small-scale variations in reddening and/or multiple dominant populations may cause the RGB to broaden, resulting in poor $|$slope$|$ estimation. For regions with lower stellar density (generally the outer regions) $N_p$ could be low, leading to poorly defined RGBs and, hence, uncertain $r$. Moreover, for such cases, the MW's contamination in the vicinity of the RGB may remain as dominant as the RGB itself, thus causing scatter in the RGB and/or uncertain slope estimation.

\begin{figure}
\centering
\includegraphics[width=\columnwidth]{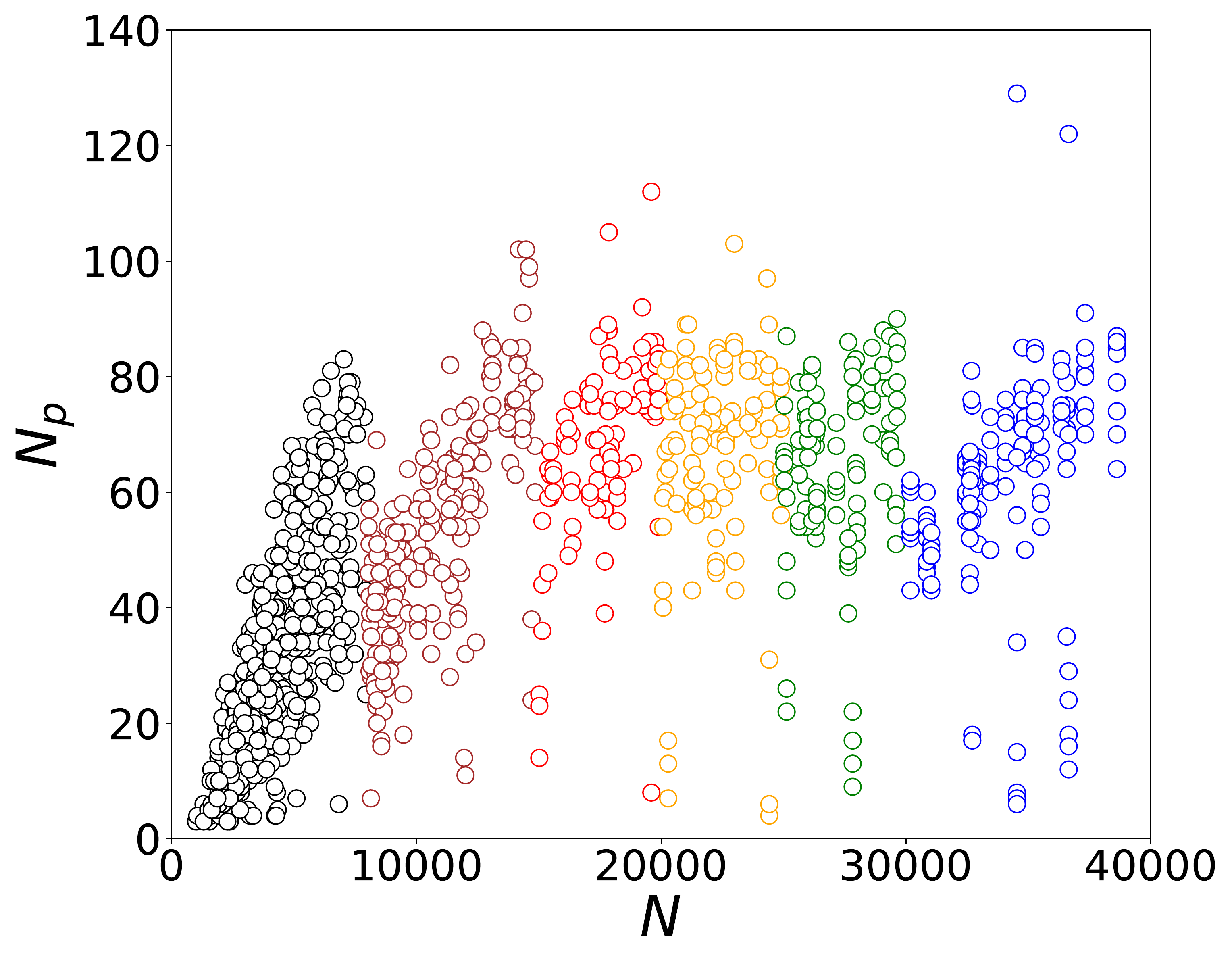}
\quad
\includegraphics[width=\columnwidth]{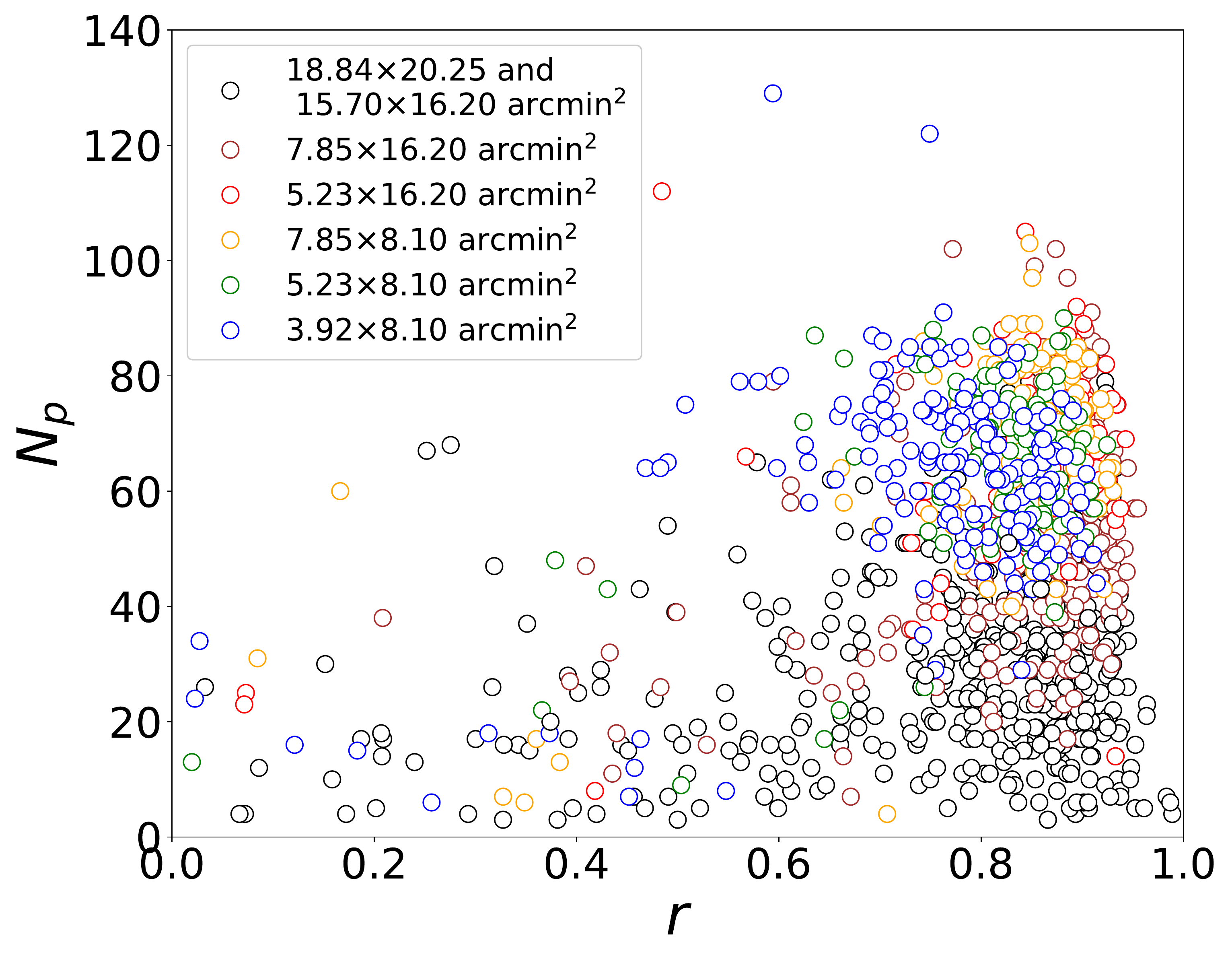}
\caption {(Top): $N_p$ versus $N$ for SMC subregions following finer binning. The colours correspond to the different bin areas; see Table \ref{table:tab1}, column (8). (Bottom): corresponding $N_p$ versus $r$ distribution.}
\label{fig:div}
\end{figure}

Table \ref{table:tab1} summarises the six division criteria used. It also lists the total number of subregions following application of each criterion, along with their corresponding areas. We adopted relatively larger areas for five tiles in the East (tiles SMC 2$\_$5, 3\_6, 4\_6, 5\_6, and 6\_5), two tiles in the North (tiles 7\_3 and 7\_4) and two tiles in the West (tiles SMC 3\_1 and 4\_1) by dividing them into 20 subregions, whereas the rest of the tiles are divided into 30 and more finer areas (we call them subregions thougout the text). We thus defined 1180 subregions, with areas ranging from 18.84$\times$20.25 arcmin$^2$ (328.75$\times$353.36 pc$^2$) to 3.92$\times$8.10 arcmin$^2$ (68.4$\times$141.34 pc$^2$). Figure \ref{fig:div} displays $N_p$ versus $N$, and $N_p$ versus $r$, in the top and bottom panels respectively. The colours of the points correspond to the six different area sizes. Thus, by adopting this method we have generated subregions with $N_p$ values confined to smaller (and similar) values for all six area-binning criteria, with $r > 0.5$.

\begin{figure}
\centering
\includegraphics[width=\columnwidth]{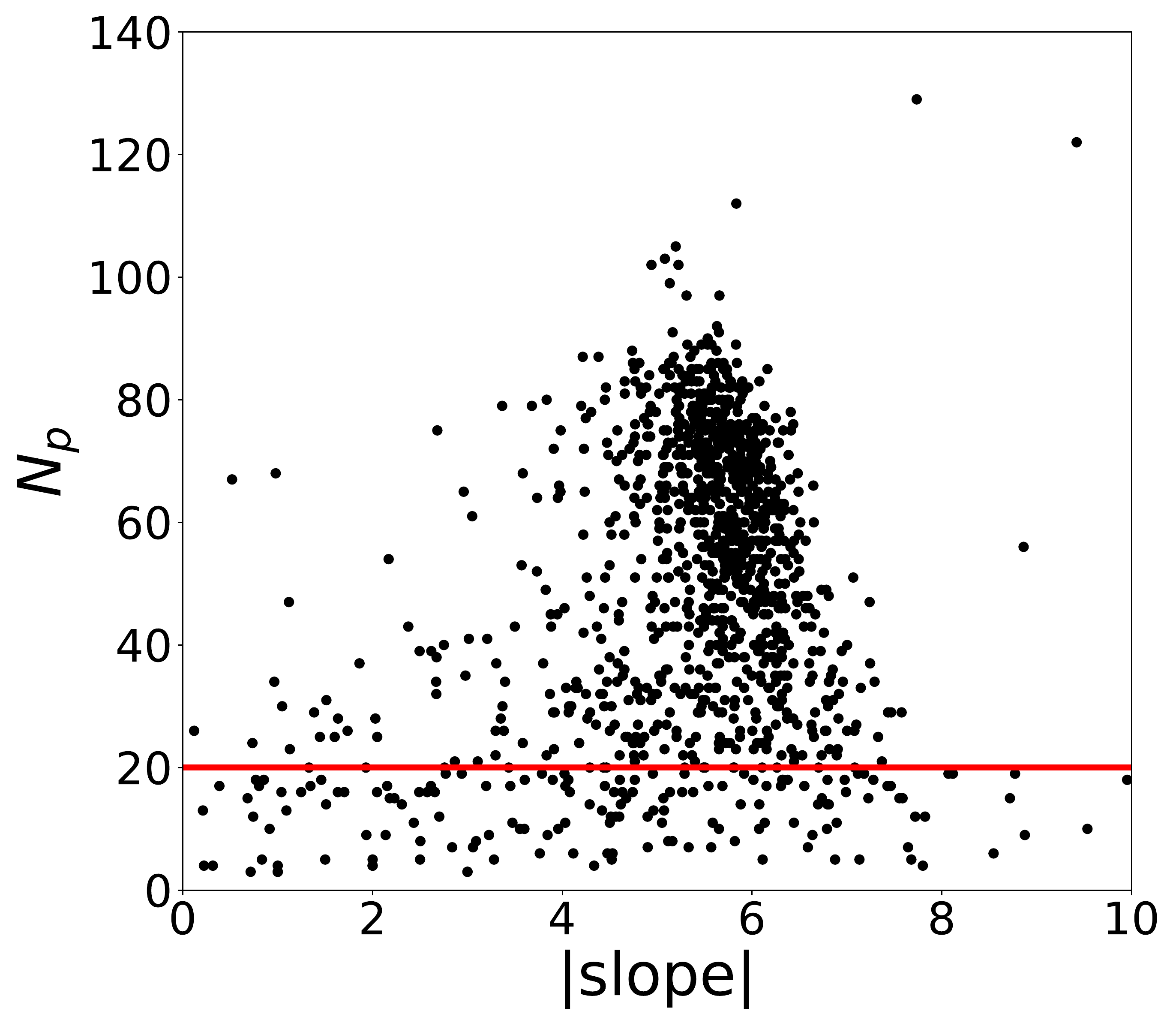}
\quad
\includegraphics[width=\columnwidth]{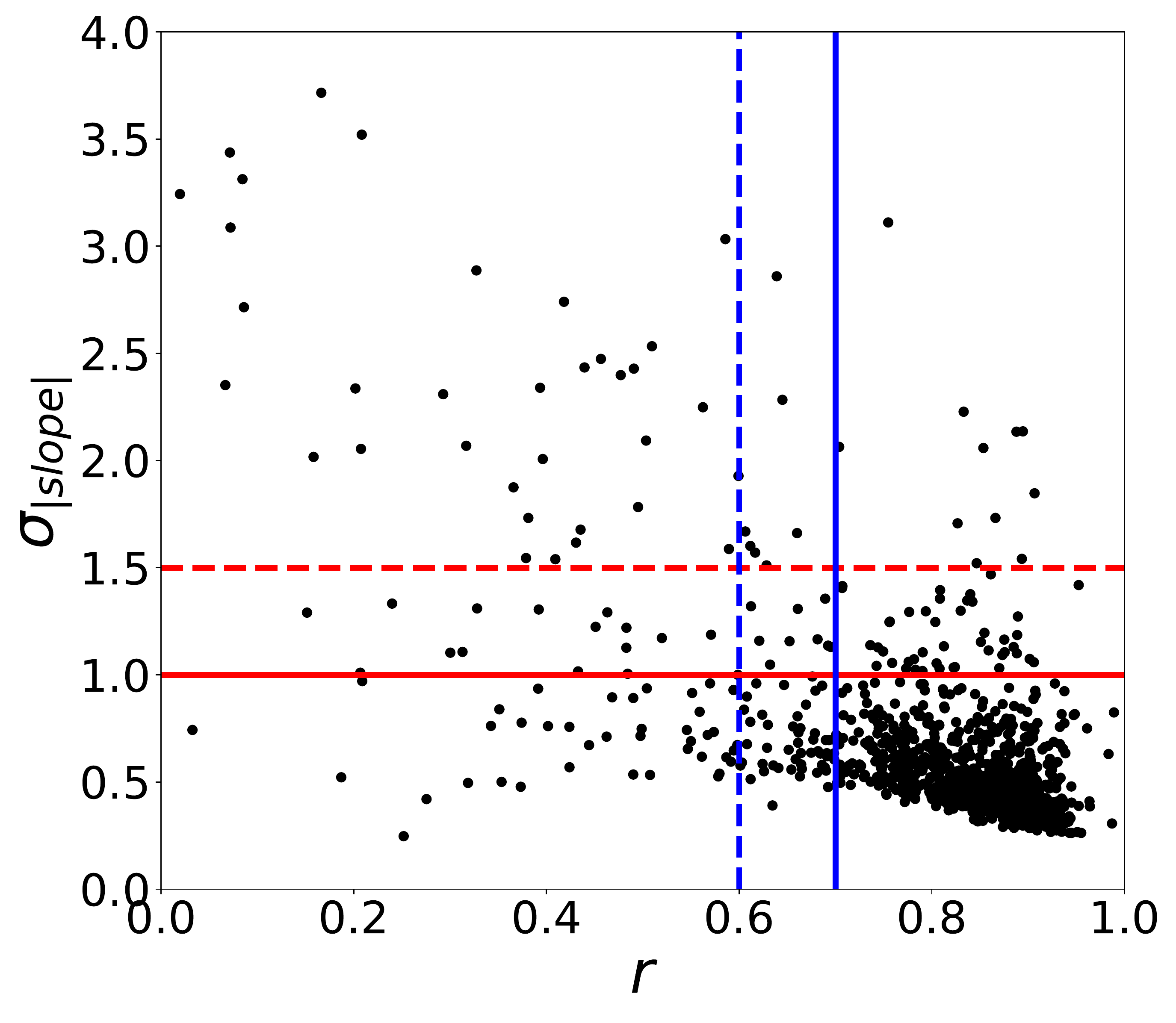}
\caption {(Top): $N_p$ versus $|$slope$|$ for the subregions. The red line at $N_p = 20$ denotes the cut-off adopted to exclude regions  with poorly populated RGBs. (Bottom): $\sigma_{\rm slope}$ versus $r$ for the subregions. The red dashed and solid lines correspond to cut-off criteria for $\sigma_{\rm slope}$ of 1.5 and 1.0, respectively. The blue dashed and solid lines denote cut-offs corresponding to $r$ = 0.6 and 0.7, respectively. } 
\label{fig:allcuts}  
\end{figure}

We exclude subregions with poorly estimated RGB slopes. In Figure \ref{fig:allcuts} (top panel), subegions with $N_p <20$ show a larger range in slope compared with regions with other $N_p$ values. The higher and lower slope values could be artefacts owing to sparsely populated RGBs. Thus, we decided to only use those subregions with $N_p \geq 20$ stars, i.e., the fitted RGB range should contain at least 60 stars. In Figure \ref{fig:allcuts} (bottom panel) we note that most of the $r$ values lie within the range from 0.6 to 0.95, and between 1.0 and 1.5 for $\sigma_{\rm slope}$. We observe a relatively large scatter for subregions with $r < 0.6$ and $\sigma_{\rm slope} > 1.5$. In addition, the clumpiest part of the distribution is found for $\sigma_{\rm slope} < 1.0$ and $r > 0.7$. Thus, we defined four different cut-off criteria in terms of $r$ and $\sigma_{\rm slope}$, with $N_p \ge 20$, to yield well-estimated parameters.
\begin{itemize}
\item criterion (I): $r \ge 0.6$ and $\sigma_{\rm slope} \le 1.5$;
\item criterion (II): $r \ge 0.6$ and $\sigma_{\rm slope} \le 1.0$;
\item criterion (III): $r \ge 0.7$ and $\sigma_{\rm slope} \le 1.5$;
\item criterion (IV): $r \ge 0.7$ and $\sigma_{\rm slope} \le 1.0$.
\end{itemize}

\begin{figure} 
\includegraphics[width=\columnwidth]{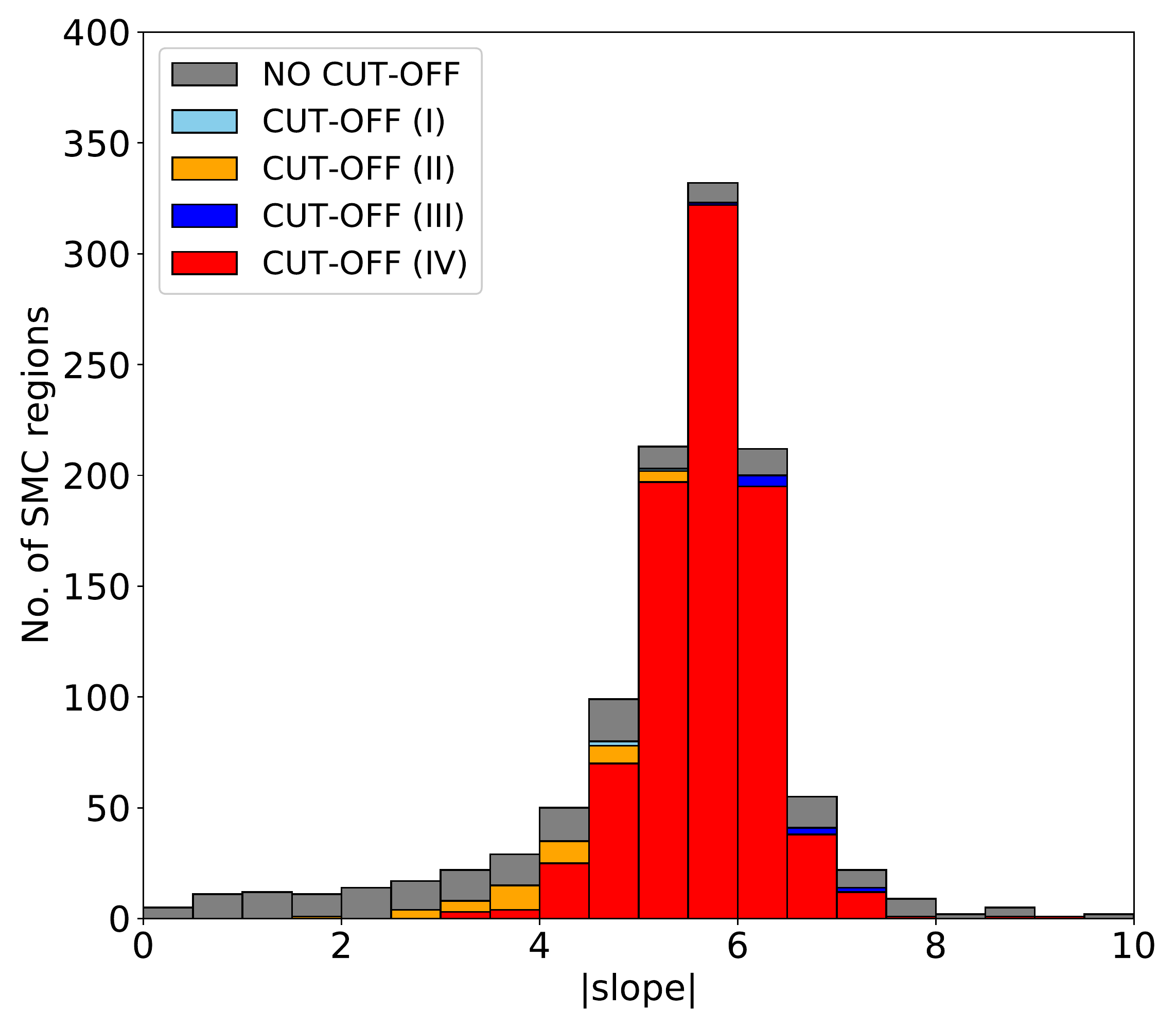}
\caption{$|$slope$|$ histogram for SMC subregions pertaining to all four cut-off criteria, compared with the equivalent distribution for no cut-off.}
\label{fig:smc_histslope}  
\end{figure}

In Figure \ref{fig:smc_histslope} we compare the distribution of the slopes resulting from the application of these four cut-off criteria with the original distribution. The primary peaks appear at the same locations for all four cut-off criteria compared with the `no cut-off' case. The width of the slope distribution decreases as the cut-off criteria become increasingly stringent, i.e., from criterion (I) to (IV). The effect is better visible for smaller slope values (less than 4.0) compared with larger values. Moreover, there is hardly any change in the slope distribution if $r$ is kept constant and $\sigma_{\rm slope}$ is varied. On the other hand, the width of the distribution reduces for constant values of $\sigma_{\rm slope}$ and variable $r$.

Note that the tile SMC 5\_2 is dominated by the Galactic globular cluster 47 Tucanae \citep[see for e.g.][]{Li+2014ApJ}. We checked a few CMDs and found that our technique cannot be used to estimate the RGB slope of the underlying SMC field. Such subregions are automatically removed by our cut-off criterion (I) and additionaly by criterion (IV), which reflects the efficacy of our adopted cut-off criteria. We inspected the remaining CMDs and removed two additional subregions manually.

\begin{table*}
\caption{Subdivision of the SMC area: The table describes the six binning criteria used to subdivide SMC tiles. For each criterion, column (2) indicates the limit on the total number of stars ($N$) within a region. Column (3) lists the number of regions within that limit. Columns (4) and (5) specify the number by which regions are binned along RA and Dec, respectively. Column (6) lists the total number of subregions. Column (7) gives the area of each such subregion, and column (8) denotes the total number of subregions corresponding to each of the six subdivision criteria. The colours adjacent to the numbers are used to denote them in Figure \ref{fig:div}'s top and bottom panels.}
\label{table:tab1}
\begin{tabular}{|c|c|c|c|c|c|c|c|}
\hline \hline
        & Stars   & Regions  & RA         & Dec        & No. of            & Area         &  Subregions    \\
        &         &          & divisions  & divisions  & divisions         & (arcmin$^2$) &  ($a \times d$) \\
        &         & (a)      & (b)        & (c)        & $(d= b \times c)$ &              &                 \\
\hline\hline
1  & 0 $<$ $N$ $\le$ 8000      & 497 & 1 & 1 & 1  & (18.84$\times$20.25) & 497 (black)     \\
   &         &  &  &  &   &  and (15.70$\times$16.20) & \\
2  & 8000 $<$ $N$ $\le$ 15,000   & 103 & 2 & 1 & 2  & (7.85$\times$16.20) & 206 (brown)    \\
3  & 15,000 $<$ $N$ $\le$ 20,000   & 32 & 3 & 1 & 3  & (5.23$\times$16.20) & 96 (red)       \\
4  & 20,000 $<$ $N$ $\le$ 25,000  & 29 & 2 & 2 & 4  & (7.85$\times$8.10) & 112 (orange)    \\
5  & 25,000 $<$ $N$ $\le$ 30,000 & 18  & 3 & 2 & 6  & (5.23$\times$8.10) & 108 (green)    \\
6  & $N$ $>$ 30,000             & 20  & 4 & 2 & 8  & (3.92$\times$8.10) & 160 (blue)\\
\hline     
\end{tabular}
\vskip 1.0ex
\end{table*} 

\begin{figure} 
\includegraphics[height=0.75\columnwidth,width=\columnwidth]{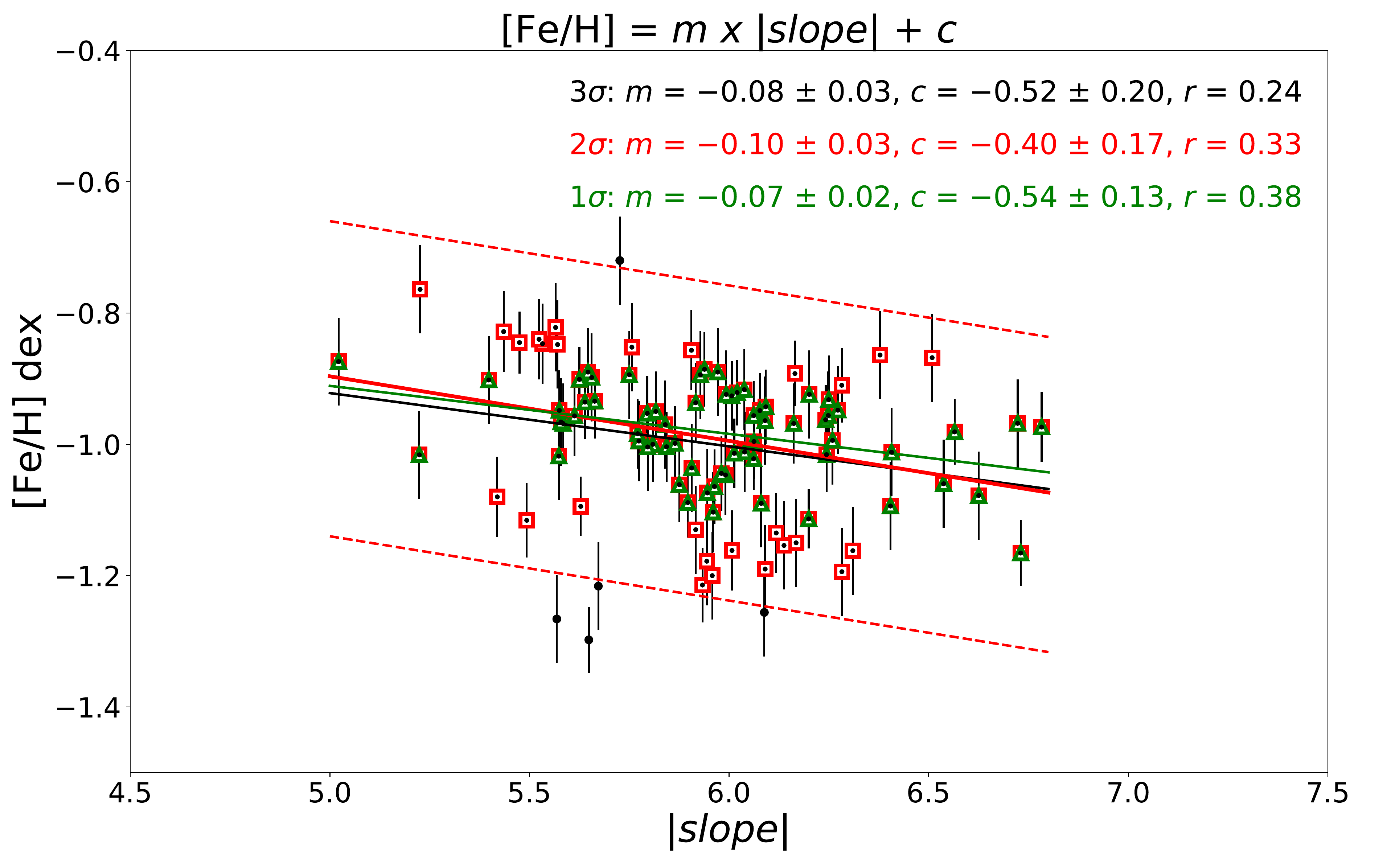}
\caption{[Fe/H] versus $|$slope$|$ for the SMC. The points denote our subregions with mean [Fe/H] values based on RGs from D14. Black points and black solid line denotes the 3$\sigma$ clipped points and their corresponding fit. The error bar (black vertical lines) shown for each point is the standard error on the mean [Fe/H].  The red open squares and red solid line denote the 2$\sigma$ clipped points and their corresponding best fit. The red dashed lines show the limits ($\pm$2$\sigma$) associated with the red solid line. This line was adopted for our slope--metallicity calibration. The green open triangles and green solid line represent 1$\sigma$ clipped points and their corresponding best fit.}
\label{fig:smc_calib1}
\end{figure}

\subsection{Calibration of the RGB slope as a function of metallicity}

Once we had estimated the RGB slope values for all subregions, our next step was to calibrate these values with respect to metallicity. To derive such a calibration relation we need spectroscopic studies of the same stellar population (RGs) with a range in metallicity values, covering a large spatial area similar to the VMC survey. We used the spectroscopic study of D14. 
Following C18, we estimated mean metallicities of subregions in the SMC by averaging over the D14 RGs located within the relevant subregion, and plotted them against their corresponding RGB slope values (Figure \ref{fig:smc_calib1}). We considered stars located within twice the standard deviation about the mean metallicity for subregions that contained at least 5 RGs (see also C18). To ensure a reliable calibration relation, we used cut-off criterion (IV) to select subregions with relatively better RGB slope estimates, and we further constrained this cut-off by considering subregions with $\sigma_{\rm slope} \leq 0.50$ (rather than $\leq 1.0$). We then visually inspected individual CMDs to ensure that we selected subregions with best-fitting RGB for calibration by not taking into account any artefacts or spurious cases. 

Figure \ref{fig:smc_calib1} shows the resulting mean [Fe/H] versus RGB slope relation for 97 subregions. The RGB slopes range from 5.0 to 6.8, covering most of the RGB slope distribution (Figure \ref{fig:smc_histslope}), whereas the metallicity range covered the range from $-0.70$ to $-1.30$ dex.  
To estimate a slope--metallicity relation, we performed a linear least-squares fit with three different clipping choices (3$\sigma$, 2$\sigma$ and 1$\sigma$). The slope and y-intercept values for all three agree within the errors. The 1$\sigma$ clipped relation, although it has the highest $r$, results in a lower reliability of the estimated slope--metallicity relation. We used the 2$\sigma$ clipped relation owing to a higher $r$ as compared to 3$\sigma$ case. Our calibration relation is:
\begin{equation} \label{eq:1} 
{\rm [Fe/H]}= (-0.10\pm0.03)\times|{\rm slope}|+ (-0.40 \pm 0.17) dex,
\end{equation}
with $r=0.33$. 
The associated probability of the correlation, $p = 0.0012$ (significant at the level of $P < 0.05$). The 1$\sigma$ dispersion of the residuals between the observed metallicity and the best-fitting line is 0.10 dex. Note that our calibration of the RGB slope to metallicity rests on the assumption that the spectroscopic targets are drawn from the dominant population of the subregions. 
C18 derived slopes of the calibration relation of $-0.10 \pm 0.03$ dex for OGLE III and $-0.08 \pm 0.02$ dex for the MCPS using $V$ vs $V-I$ CMDs, which agrees within the uncertainities with the value derived here. We observe a scatter resulting in lower $r$ compared with the slope--metallicity calibration relations of C18 ($r \sim 0.50$). However, using the VMC data, we were able to use about twice more calibration points than C18, and thus our statistical basis is improved.

\begin{figure} 
\includegraphics[width=\columnwidth]{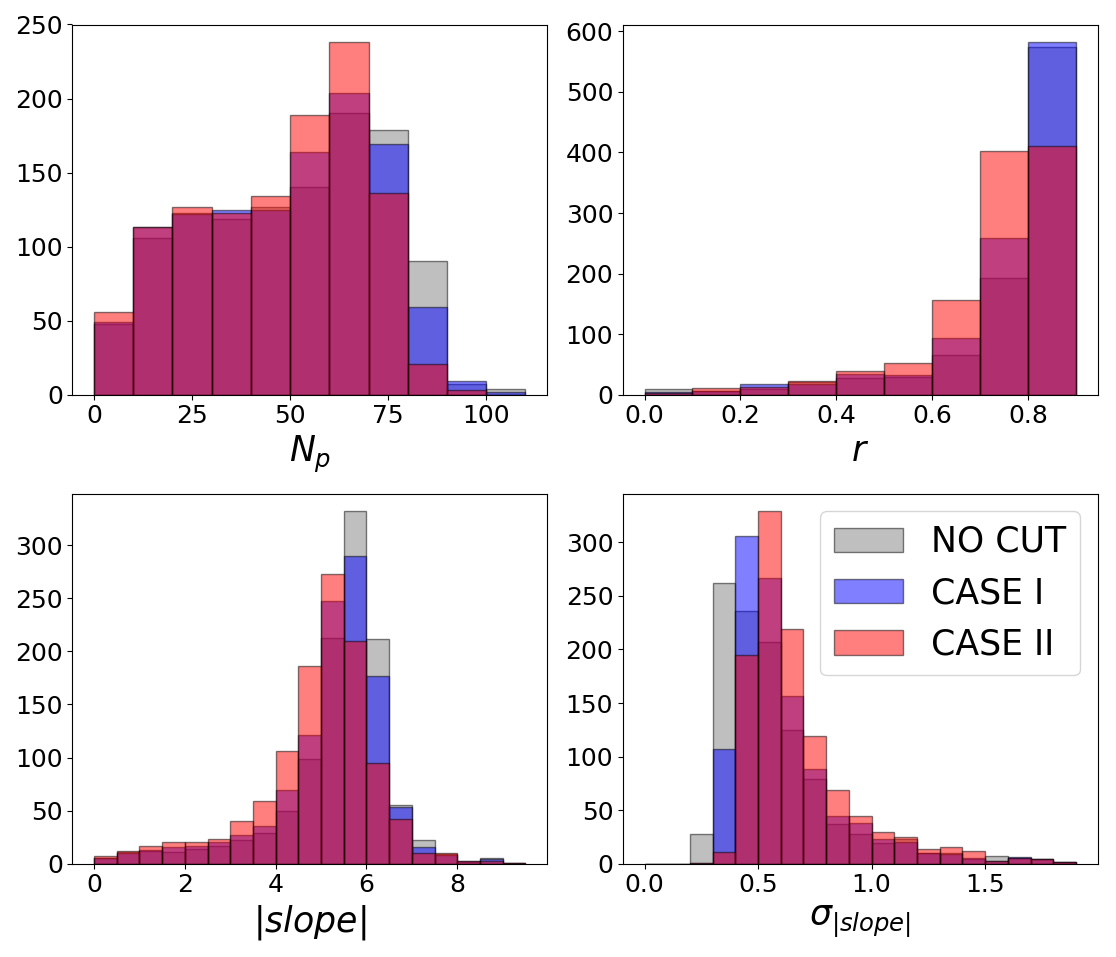}
\caption{Histograms of parameters estimated for the RGB with no
  restriction on the brightest end (grey), RC peak $-$3.0 mag (blue), and RC peak $-$2.5 mag (red). The top left-hand panel shows a comparison between the three cases for estimates of $N_p$, $r$ at the top right and $|$slope$|$ at the bottom left. The bottom right shows the $\sigma_{\rm slope}$ histogram.}
\label{fig:para_rc1rc2}
\end{figure} 
 
The dispersion in values of metallicity for similar values of RGB slope could be owing to our selection of RGB stars in the subregions (systematic) or to real (astrophysical) reasons. One factor that could have affected our RGB slope estimates is the extent of the RGB. We adopted the peak of the RC as the base of the RGB. Although the actual base could be different, we consistently selected the same faintest boundary all subregions. Selection of a minimum number of stars ($\geq$ 3) within the RGB bin restricts the brightest end of the RGB from the RC peak to about 3 mag brighter, or even more for some regions. If the brightest end of the RGB is not fixed, this can have an effect on the RGB slope estimates for different subregions (in view of the curvature of the upper magnitude range), thus leading to systematic uncertainties. To inspect the importance of this effect on the slope--metallicity relation, we estimated the RGB slopes for all subregions by considering two extents of the RGB. For Case I, we constrained the brightest extremity to 3.0 mag brighter than the RC peak magnitude, and for Case II, we restricted it to 2.5 mag brighter. We then estimated the parameters for these two cases and compared them with those obtained without having imposed any constraints. Figure \ref{fig:para_rc1rc2} shows a comparison of the estimated parameters. The widths of the $|$slope$|$ distribution for all three cases are very similar, with the peak of the distribution for Case II shifting slightly towards lower slope values. A comparison of the $r$ values shows that for Case II, $r$ has a broader peak compared with the other two cases. The widths of $\sigma_{\rm slope}$ are also similar for all three cases. However, the peak values shift towards larger errors for Cases I and II, with a relatively larger shift for Case II.

Next, we checked for effects on the slope--metallicity relations for both cases. We found that the scatter in the slope--metallicity plane for Cases I and II was unaffected. The estimated slope--metallicity relations are similar to that estimated above with no improvement in $r$. Thus, constraining the brightest end (or the extent of the RGB) does not have a significant effect on the slope--metallicity relation derived. This could mean that the scatter in the plot could be real, reflecting that the dependence of the RGB slope on metallicity in the SMC is marginal for this RGB slope range, or a larger dispersion in metallicity values for a given RGB slope. The latter could mean that RGB stars of different metallicities are well-mixed spatially.

In summary, we find that the $|$slope$|$ distribution is similar for all three cases, with larger $\sigma_{\rm slope}$ for Cases I and II. The three slope--metallicity relations are similar, with the no-constraint case having a larger number of data points and a larger value of $r$ compared with Cases I and II. Since we are more interested in the relative change in slope values (and, hence, metallicities) among subregions, this choice will not affect our results.


\section{NIR metallicity maps of the SMC}
 
\begin{figure*} 
\begin{center} 
\includegraphics[height=3in,width=6in]{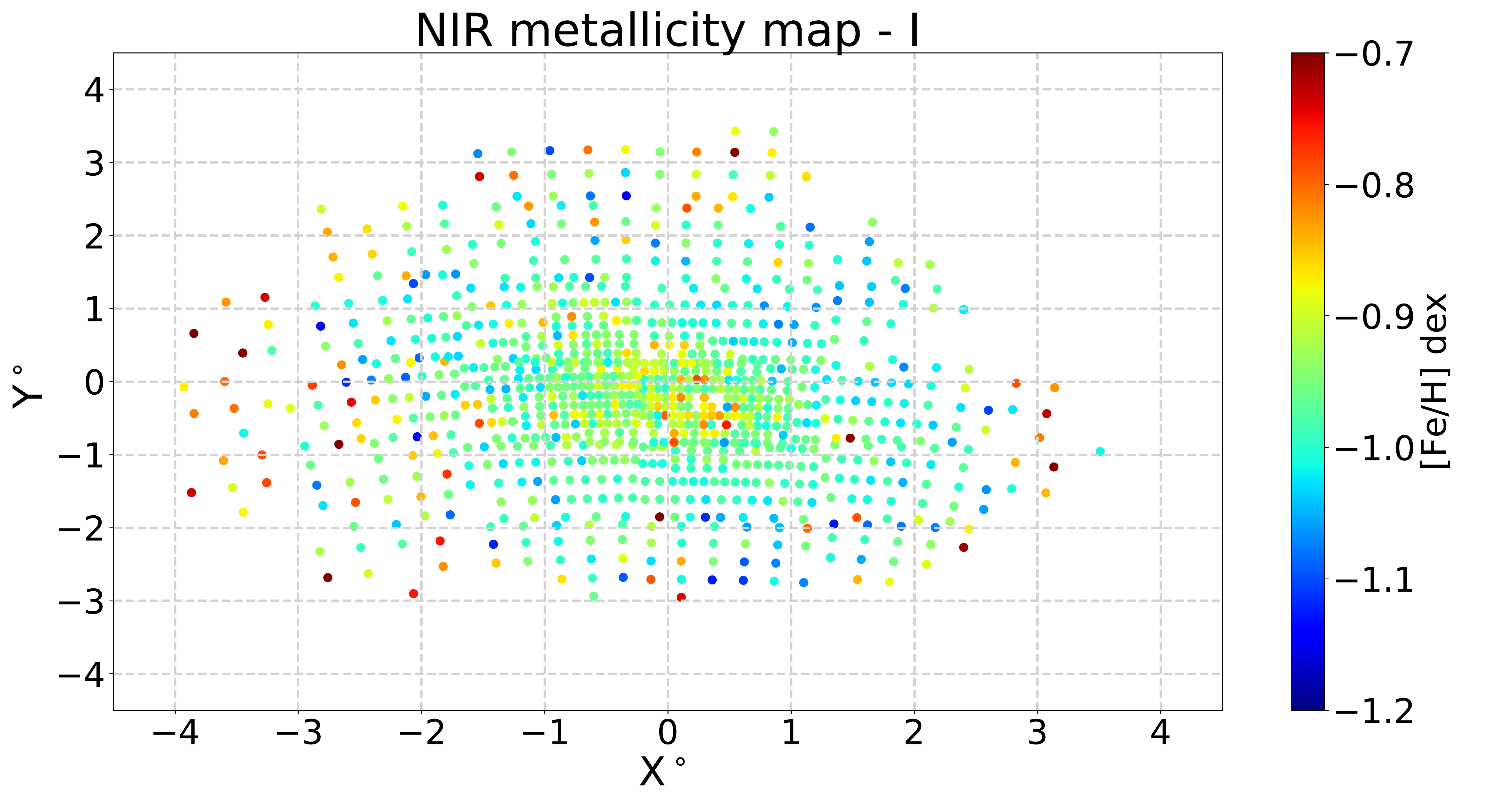}
\caption{Metallicity map based on cut-off criterion (I): $N_p \ge 20,
  r \ge 0.6$ and $\sigma_{\rm slope} \le 1.5$.}
\label{fig:nirmap1}
\end{center} 
\end{figure*}

\begin{figure*} 
\begin{center} 
\includegraphics[height=3in,width=6in]{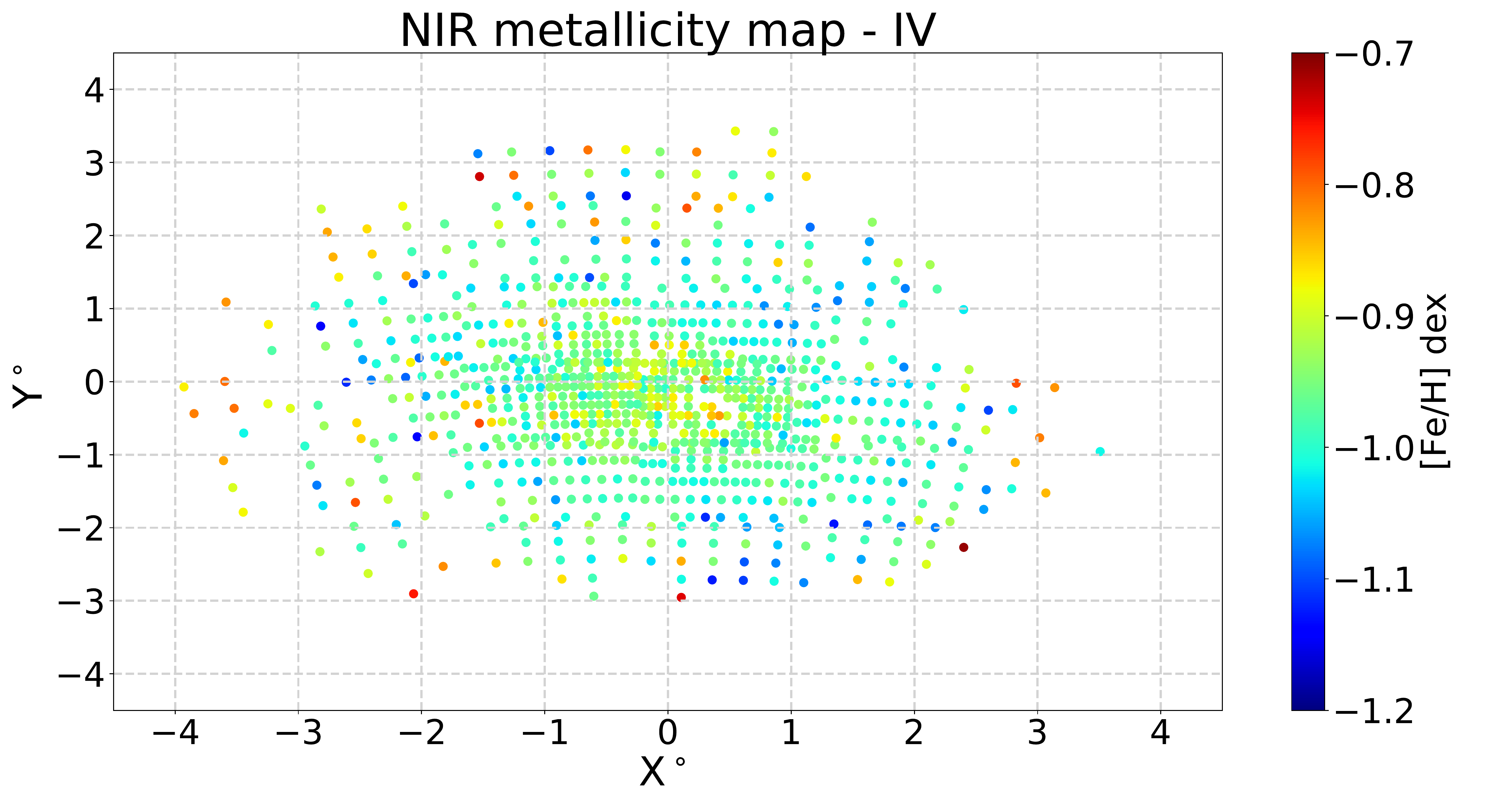}
\caption{Metallicity map based on cut-off criterion (IV): $N_p \ge 20,
  r \ge 0.7$ and $\sigma_{\rm slope} \le 1.0$.}
\label{fig:nirmap4} 
\end{center} 
\end{figure*}

Our next step was to convert all estimated RGB slopes to metallicities using Equation \ref{eq:1}. To understand the spatial metallicity variation in the plane of the sky we convert RA and Dec to the Cartesian $(X, Y)$ plane. For the SMC's centre we adopted RA = 0$^{\rm h}$ 52$^{\rm m}$ 12.5$^{\rm s}$ and Dec = $-$72$^{\circ}$ 49$'$ 43$''$ \citep[J2000.0][]{deVaucouleurs&Freeman1972VAstructure}. The resulting VMC metallicty maps for the SMC are shown in Figures \ref{fig:nirmap1}
and \ref{fig:nirmap4} for cut-off criteria (I) (935 subregions) and (IV) (875 subregions), respectively. Since the resulting maps for cut-off criteria (II) and (III) appear very similar to each other and do not add much to our results, we present only the two extreme cases to highlight their differences.

Metallicity trends in the central SMC, out to a radius of about 3--4${^\circ}$ towards the East and West, and also out to 3${^\circ}$ to the North and South are revealed. The area covered is about two to three times larger than that covered by C18's maps using OGLE III and MCPS data (their figures 11 and 22, respectively). Overall, both NIR maps are dominated by subregions corresponding to a metallicity range from $-1.10$ to $-0.80$ dex. 
The region around the centre of the SMC (within 1--1.5${^\circ}$) is relatively metal-rich ($>$ $-0.85$ dex) with respect to rest of the field but not homogeneous. This is similar to the results of C18. The NIR maps show some scattered metal-rich points in the outer regions of the SMC (beyond 3${^\circ}$ East--towards the Wing, 2.5${^\circ}$ North and 3.0${^\circ}$ West). Since the C18 maps are limited by the survey coverage of OGLE III and MCPS, metallicity trends for subregions beyond 2.5${^\circ}$ in radius could not be compared with the NIR map. 

Applying the more stringent criteria (I to IV) removes subregions (18 to 25 per cent of total) located around the SMC's centre and in the outer regions beyond 2.5${^\circ}$--3.0${^\circ}$ (mainly towards the East -- Wing and North, and a few towards the West). Our NIR metallicity maps appear more complete within the central 2.5${^\circ}$--3.0${^\circ}$ compared with the range from 3.0${^\circ}$--4.0${^\circ}$ towards the East, West and North. In fact we have been able to estimate metallicity trends for a larger area within 1--1.5${^\circ}$ of the SMC's centre than C18. The gaps observed are subregions that suffer from differential reddening (near the centre and the eastern Wing), host extreme mixtures of stellar populations, and/or are affected by variations in the LOS depth (mainly near the centre and the eastern part of the SMC). 
There could also be an effect of crowding in the central region, leading to poor RGB slope estimation, hence contributing to gaps. This requires further investigation which is beyond the scope of this work. The effects of reddening and variation in LOS depth are discussed in Sections 5.1.1 and 5.1.2, respectively.

\begin{figure}
\includegraphics[width=\columnwidth]{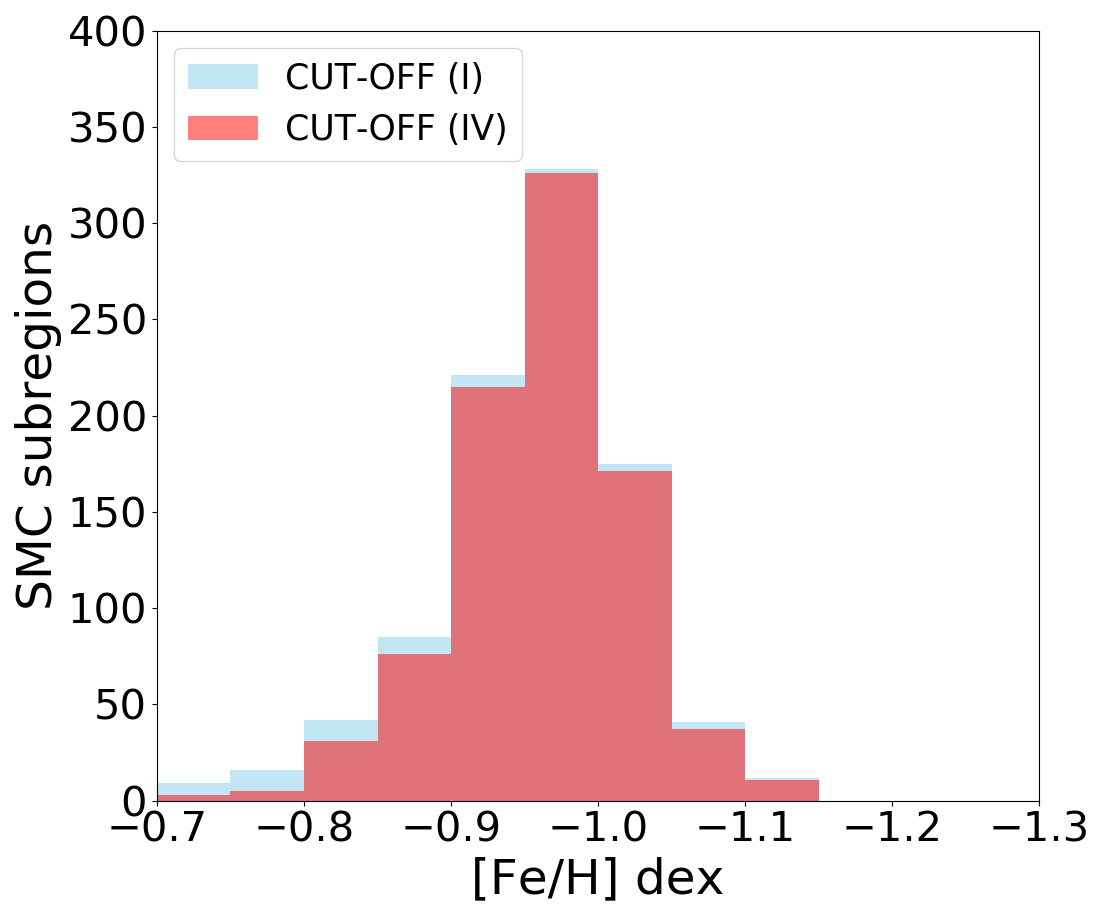}
\caption{Histogram of [Fe/H] estimated for cut-off criteria (I) (blue) and (IV) (red).}
\label{fig:histabun_allcuts}
\end{figure}

\begin{figure}
\includegraphics[width=\columnwidth]{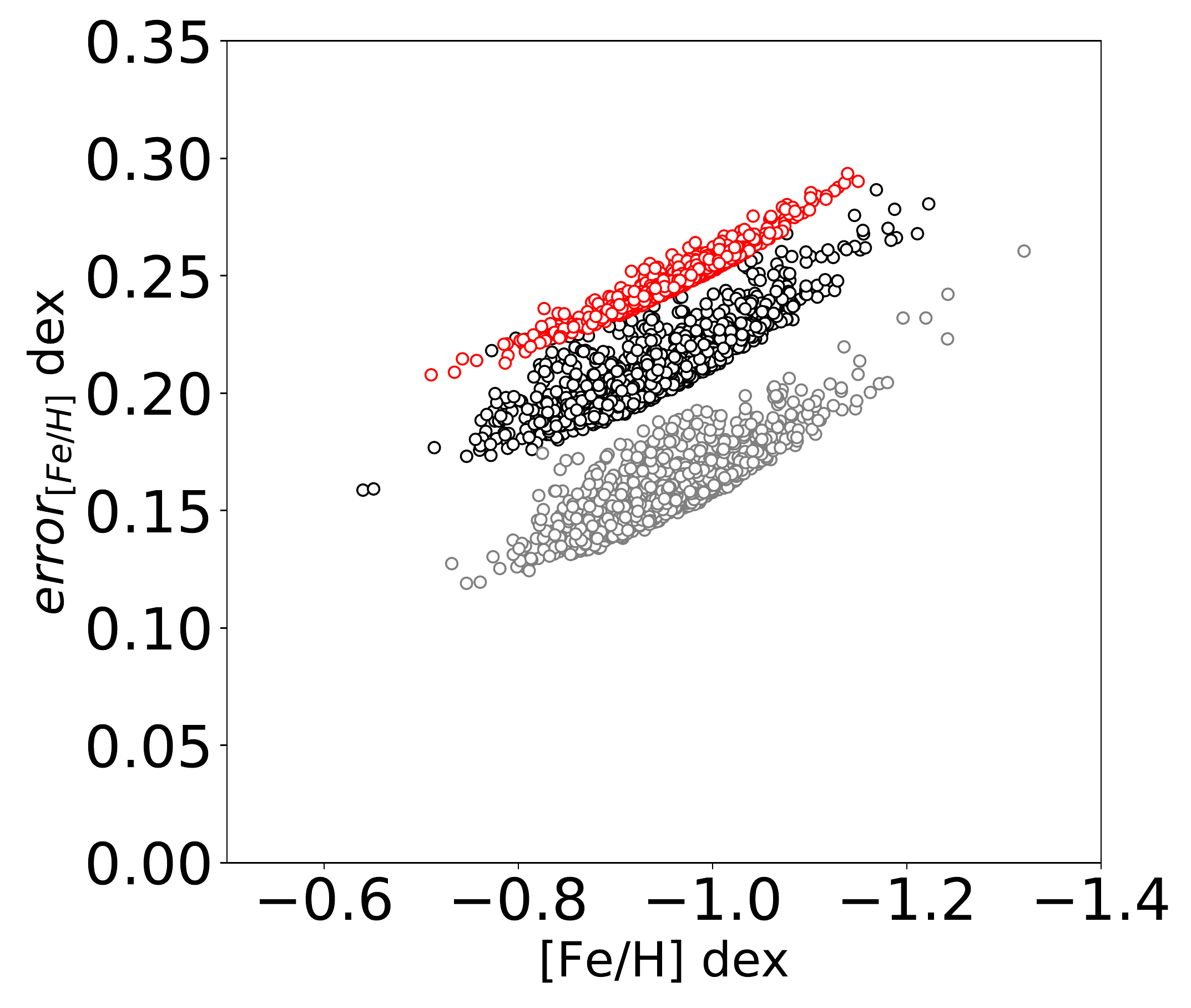}
\caption{error$_{\rm [Fe/H]}$ versus [Fe/H]. VMC: red; MCPS: grey
  (C18); OGLE III: black (C18).}
\label{fig:error} 
\end{figure}

The metallicity histograms pertaining to criteria (I) and (IV) are shown in Figure \ref{fig:histabun_allcuts} for comparison. Both distributions are similar to a single peak around $-0.975$ dex. The mean metallicities of the SMC for criteria (I) and (IV) are : $-0.95$ $\pm$0.07 dex and $-0.96$ $\pm$0.06 dex, respectively. The errors are the standard deviations of the mean and do not include the error associated with the metallicity estimation of each subregion. The mean values for both criteria are almost identical, although the number of SMC subregions is different (935 versus 875). We have calculated the errors associated with our metallicity estimates (${\rm error}_{\rm [Fe/H]}$) using error propagation, following Equation 3 of C18. For a detailed explanation of the error estimation we direct the reader to their Section 5. Figure \ref{fig:error} shows a comparison of the error values with C18, i.e., for the OGLE III and MCPS data sets (black and grey points, respectively). The range of values spans from $\sim$0.20 to 0.30 dex, i.e., larger than the OGLE III and MCPS errors. The dispersion in error$_{\rm [Fe/H]}$ ($\pm$ 0.01 dex) for a given value of [Fe/H] is determined by the range of $\sigma_{\rm slope}$ associated with the corresponding RGB $|$slope$|$. There is a slight trend suggesting that our estimated errors are higher for metal-poor values. The estimated error$_{\rm [Fe/H]}$ for the VMC data are larger than for the MCPS and OGLE III data, for the same [Fe/H]. This is most likely related to the difference in the corresponding values of $|$slope$|$ and the slope--metallicity calibration relation, which are functions of the wavelengths used in these studies.



\section{Discussion}

\begin{figure}
\includegraphics[width=\columnwidth]{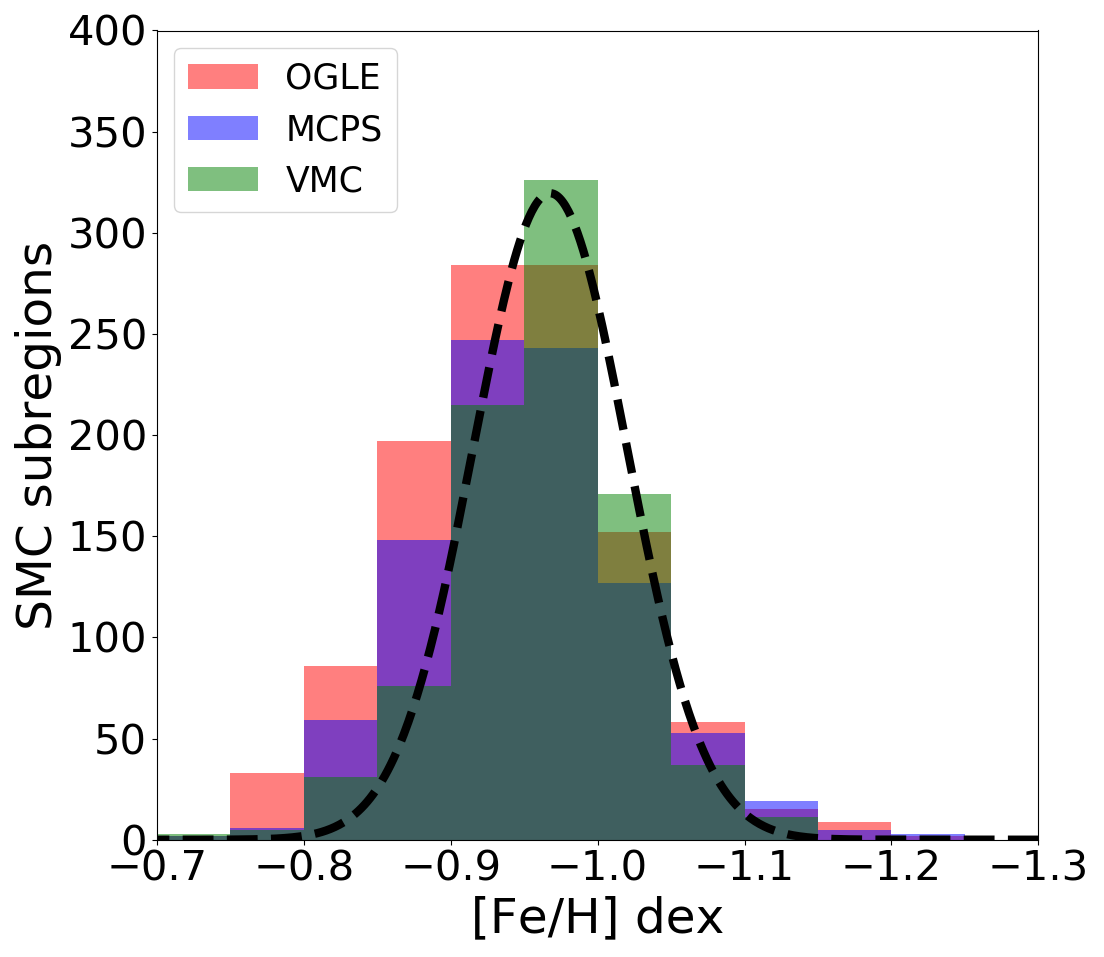}
\caption{Metallicity histogram for the entire SMC using VMC data fitted with a Gaussian function (black dashed) for cut-off
  criterio (IV). The corresponding distributions for the OGLE III and MCPS surveys are also shown for comparison (C18).}
\label{fig:histabun}
\end{figure}
\subsection{Impact of reddening and LOS depth}

\subsubsection{Effects of reddening and differential reddening}

\cite{Haschke+2011} used RC stars and RR Lyrae from OGLE III survey to study the reddening across the SMC. These authors reported a mean SMC reddening of $E(V-I)= 0.04 \pm 0.06$ mag using RC stars and a median reddening of $E(V-I) = 0.07 \pm 0.06$ mag based on RR Lyrae data. Their SMC reddening maps (their figures 4 and 11) show two locations along the bar exhibiting higher reddening than the field, while the highest reddening is found in the star-forming Wing. The SMC reddening map of \cite{Smitha&Purni2012}, using RC stars, shows less spatial variation across the galaxy. In addition, the southwestern and northeastern regions around the SMC centre and the eastern Wing region have higher reddening than other regions in the SMC. The mean $E(V-I)$ estimated by these authors is $0.053 \pm 0.017$ mag. The recent study by \cite{Muraveva+2018} used RR Lyrae from the VMC survey, and presented an extinction map in their figure 5 which shows that the extinction is higher in the eastern/southeastern regions of the SMC. These authors found similar results as \cite{Haschke+2011} and \cite{Smitha&Purni2012}. Very recently \cite{Gorski+2020ApJ} produced revised reddening maps which show similar spatial variations of the reddening as \cite{Haschke+2011}, but their values are systematically higher.

Variations in reddening can cause the location of the RGB to shift in the CMD. In our technique, we anchor the base of the RGB to the densest part of the RC, which takes this effect into account. C18 noted poor RGB estimates in the some central and eastern (Wing) regions, possibly owing to significant variations in reddening. Use of NIR passbands offers an advantage since they suffer less from reddening effects than optical passbands. Nevertheless, effects owing to differential reddening will remain and tend to make the RGB scatter (more broadly). This can result in poor estimates of the RGB slope and hence metallicity. In our NIR metallicity maps, subregions that were eliminated from the centre and the Eastern Wing coincide with regions in the galaxy exhibiting large variations in reddening.

\subsubsection{Effects of the LOS depth}

The SMC has a substantial LOS depth. Quantification of the LOS depth depends on the methods and tracers (young, intermediate-age, old) used and ranges from a few kpc to more than 20 kpc \citep{Mathewson+1986, Mathewson+1988, Crowl+2001, Smitha&Purni2012, deGrijs+2015AJ, Smitha+2017MNRAS}. \cite{Smitha&Purni2009, Smitha&Purni2012} found the LOS depth to be almost uniform across the inner 2.5${^\circ}$ of the SMC based on analysis of the RC population using OGLE III data. \cite{Smitha+2017MNRAS} using the VMC survey suggested that the central regions of the SMC show a large LOS depth and a mild distance gradient from East to West ($-2{^\circ}$ to 2${^\circ}$), but none from North to South.

Thus, we reached consistent conclusions using our method in terms of the estimated RGB slopes and our calibration to metallicity inside a radius of 2.5${^\circ}$. The effect of a large variation of LOS depth in the eastern and northeastern SMC (beyond 2.5${^\circ}$) is that the RGB could be spread in magnitude, leading to a poor estimation of the RGB slope. Moreover, we considered larger area bins for these subregions as they are sparsely populated compared with the main body of the SMC. In such a case, we sampled a relatively larger portion of the MW as compared with the inner regions, which can affect our analysis of the RGB slope (hence metallicity). The northern and western subregions (beyond 2.5--3${^\circ}$) do not exhibit variations in LOS depth as in the East, but sparse RGB and MW contamination (redder feature) can affect their RGB slope estimation. The impact of the subregions that remain (on the estimated MG) is discussed in Section 5.3. 

Studies suggest that the eastern and northeastern regions of the SMC are distorted by LMC--SMC interactions. A stellar structure in the eastern region closer than the SMC's main body ($d \sim 55$ kpc), at 4${^\circ}$ from the SMC's centre, was identified by \cite{Nidever+2013}. \cite{Smitha+2017MNRAS} using 13 tiles from the VMC survey found a tidally stripped RC population in the eastern regions of the SMC. This population is located at 2.5--4${^\circ}$ East of the SMC's centre, $\sim 11.8 \pm 2.0$ kpc in front of the galaxy's main body. Their figure 6 shows the luminosity function of the two RC populations. The RC population lying at the distance of the SMC main body is dominant compared with the nearer RC population, except for tiles SMC 5$\_$6 and 6\_5 which are located towards the North-East. Tatton et al. (in preperation) analysed RC stars using all 27 tiles from the VMC survey. They suggest the existence of RC bimodality in three additional eastern tiles (SMC 2$\_$5, 3$\_$6, 4$\_$6) in addition to tiles SMC 5$\_$6 and 6\_5. Similarly to \cite{Smitha+2017MNRAS}, their study (figure 6) also suggests the dominance of the RC lying at the SMC's distance over the nearer component, except for the eastern tiles. Since our technique identifies the most dominant population, it will select the RC at the SMC's distance as the RGB's base, which is thus consistent with the majority of the other tiles. For tiles SMC 6$\_$5, 5$\_$6 and 4$\_$6 the nearer RC peak dominates. Meanwhile, both RC components have similar RC peaks for tiles SMC 2$\_$5 and 3$\_$6. For subregions within these five eastern tiles, it is possible that our technique may lead to inconsistent determination of the RC peak (RGB base) which might lead to sizeable errors in RGB slope and/or poor $r$. These sugregions were automatically discarded by our cut-off criteria, thus contributing towards missing subregions towards the East/North-East in our NIR metallicity maps.
 
\subsection {Mean metallicity of the SMC}
 
The area sizes of the subregions adopted in this study are different from those of C18. This is because all three surveys are characterised by different photometric depths and angular resolutions. Moreover, we used NIR passbands to trace the RGB, whereas C18 used optical passbands. The VMC subregions are 2--3 times larger in area than the equivalent MCPS and OGLE III subregions at a specific location. Although we have calibrated all three studies onto the same metallicity scale, it is difficult to consider one-to-one matches between metallicities estimated for each subregion. This is because, as the area of the subregion varies, the dominant population and differential reddening may also vary, causing the RGB slope (and, hence, the metallicity) to vary as well (within the uncertainties). Thus, in this section and in Section 5.3 we only compare our results for the global mean and spatial variations using VMC data with the equivalent parameters pertaining to OGLE III and MPCS.
 
Figure \ref{fig:histabun} shows the SMC's metallicity distribution as estimated using cut-off criterion (IV) compared with C18. The metallicity values for all three data sets peak at similar values. A Gaussian approximation to the VMC metallicity distribution suggests a mean metallicity of [Fe/H] $= -0.97 (\sigma$[Fe/H] = 0.05) dex. A comparison with the values estimated by C18 for OGLE III ([Fe/H] = $-0.94, \sigma$[Fe/H] = 0.09 dex) and MCPS data ([Fe/H] = $-0.95, \sigma$[Fe/H] = 0.08 dex) suggests that neither the difference in area coverage of the three surveys nor the difference in the typical areas of the subregions considered in each of the three cases have a significant effect on the mean metallicity estimation, as they agree within $1\sigma$ uncertainties. A prominent difference in the distribution is observed for [Fe/H] $> -0.95$ dex where the count of optical subregions is larger compared with the VMC subregions. This could be because the metal-rich subregions ($> -0.80$ dex) observed near the SMC's centre in OGLE III and MCPS are missing from the VMC maps. We suspect that this is owing to our selection against young ($<$ 2--3 Gyr) and metal-rich RGB populations (see Section 5.3). 
The VMC survey not only covers the OGLE III and MCPS survey areas, but its overall coverage extends well beyond these optical surveys in all four directions. Therefore, here we have been able to estimate the global mean metallicity of the SMC based on a more spatially extended data set. The peak of our metallicity distribution agrees well with the current peak stellar metallicities of the SMC, [Fe/H]= $-$1.0 dex \citep{Nidever+2019ApJ, Nidever+2020ApJ}.

The mean metallicity estimated here agrees well with the spectroscopic (CaT of RGs) estimates of \cite{Carrera+2008AJ-CEH-SMC} and \cite{Parisi+2010AJfieldI, Parisi+2016AJfieldII}. \cite{Carrera+2008AJ-CEH-SMC} derived a mean metallicity of [Fe/H] $\sim -1.0$ dex within 4$^{\circ}$ of the SMC's centre for a few hundred RGs in 13 fields. \cite{Parisi+2016AJfieldII} analysed a sample of $\sim$750 RGs (in 30 SMC fields) distributed out to about 8$^{\circ}$. A Gaussian fit to their sample (their figure 2) suggests a peak metallicity of $-0.97 \pm 0.01$ dex, with a dispersion of $0.30 \pm 0.01$ dex. The authors also combined
their sample with that of D14 to create a larger sample of RGs. A Gaussian fit to this combined sample returns a metallicity peak at $-0.976 \pm 0.007$ dex, which agrees well with our peak value ($-0.967$ dex). \cite{Parisi+2016AJfieldII} did not find any bimodality in either distribution, except for a metal-poor excess ($<$ $-$1.5 dex) which is more prominent in the combined sample. Our metallicity distribution does not show any bimodality either, but it is devoid of any such metal-poor excess. \cite{Parisi+2015AJclusII}, using a combined sample of RGs in star clusters from their previous publications, argued that there is a high probability that the metallicity distribution of RGs in clusters is bimodal with peaks at $-1.1$ and $-0.8$ dex, unlike the field-star distribution. However, their derived mean cluster metallicity of $-0.9 \pm 0.2$ dex is in good agreement with that of their field-star sample, as well as with our results presented here. 

The estimated mean SMC metallicity found here is relatively metal-rich compared with the equivalent determinations using RR Lyrae stars,
which suggest a mean metallicity $\leq -1.50$ dex \citep{Deb&Singh2010MNRAS, Haschke+2012aAJ, Kapakos&Hatz2012MNRAS}. This difference may be related to the mean age difference between the populations used in our current study (RGB stars) and that of RR Lyrae stars. Because of mass and metallicity effects on RGB evolutionary rates, it is difficult for old, metal-poor populations to be the dominant contributor to the bulk average metallicity of RGs \citep{Manning&Cole2017}.

\begin{figure}
\includegraphics[width=\columnwidth]{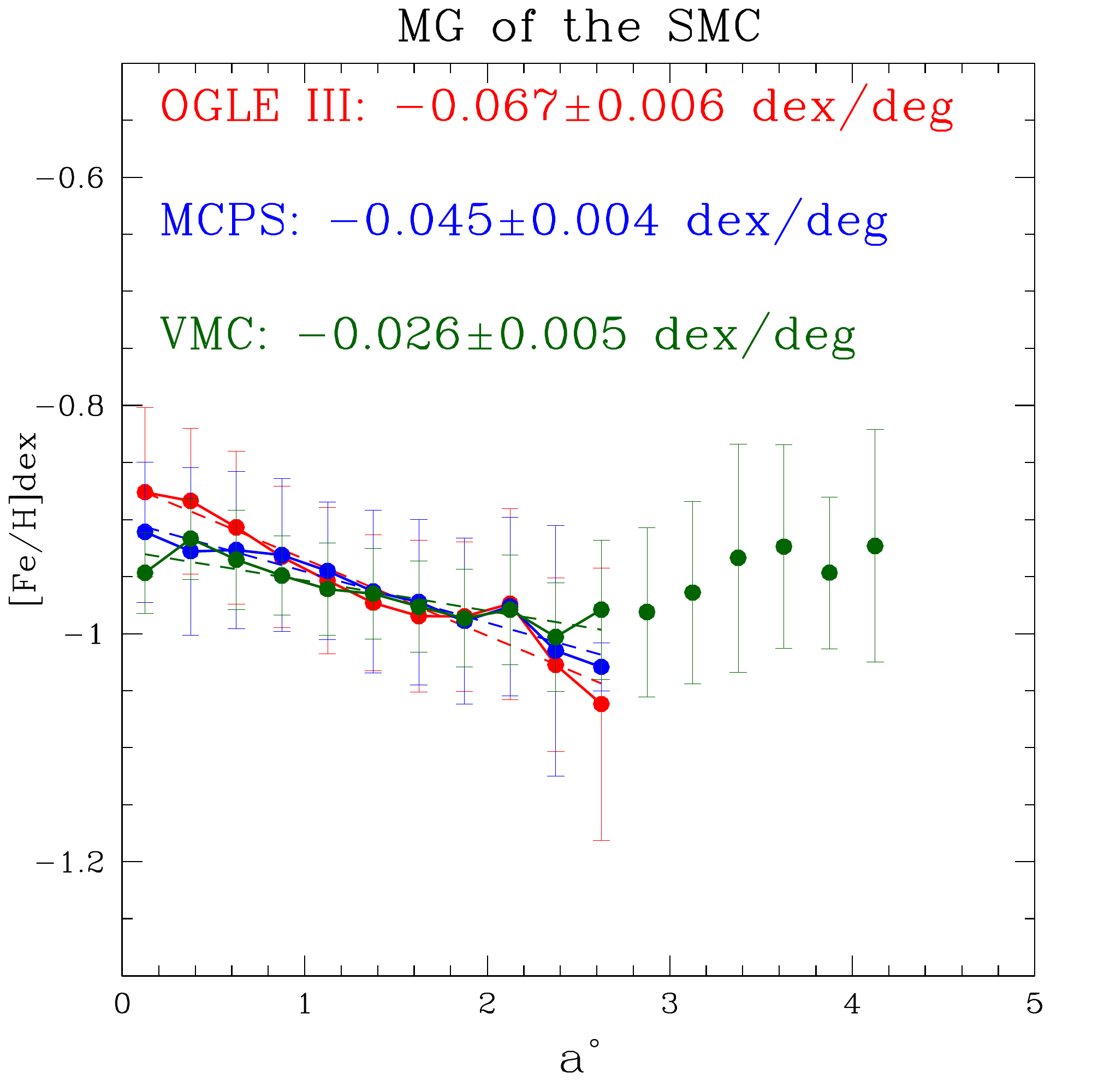}
\caption{Metallicity variation as a function of semi-major axis ($a$) of the SMC for VMC data, compared with OGLE III (red) and MCPS (blue) data, estimated for cut-off criterion (IV). The MG estimated out to $\sim 2.5{^\circ}$ is shown as a dark green line. The coloured labels show the estimated MG and the $Y$-axis intercepts for each case.}
\label{fig:metgrad} 
\end{figure}

\begin{figure}
\includegraphics[width=\columnwidth]{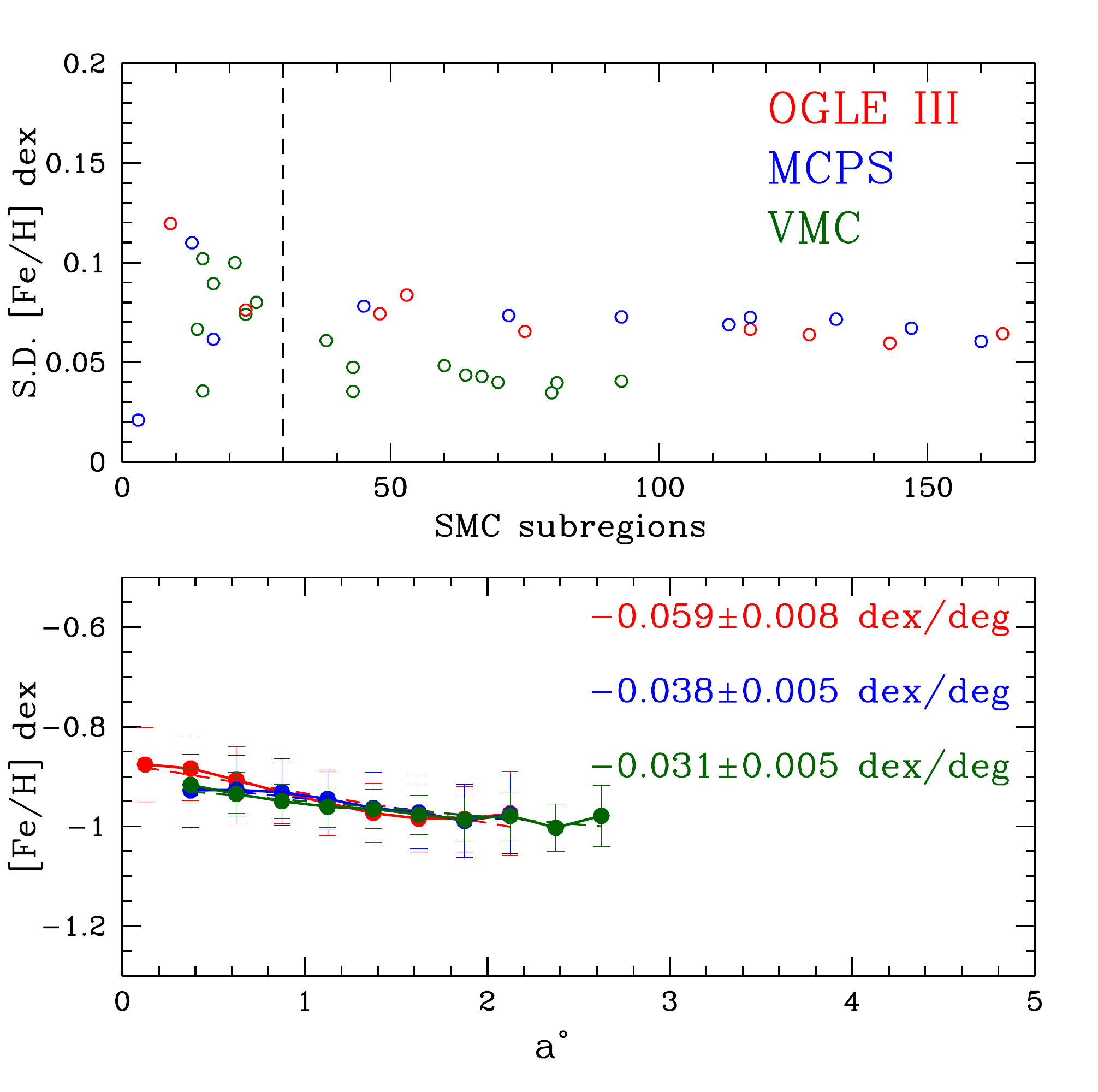}
\caption{(Top): shows the variation in the standard deviation of the mean [Fe/H] with the number of subregions within each annular bin. (Bottom): estimated MG for all  surveys for annular bins with $\ge$ 30 subregions.}
\label{fig:metgrad2} 
\end{figure}

\subsection{A metallicity gradient within the SMC}

To estimate the radial variation of metallicity within the SMC, it is important to understand orientation and projection effects. Unlike the LMC, which resembles an almost face-on disc-like system, the SMC has a more complicated structure and its orientation is poorly understood. Several studies also suggest that the SMC is markedly elongated along the LOS \citep{Gardiner&Hawkins1991MNRAS, Haschke+2012bAJ,Smitha&Purni2012,JD+2016AcA}. This renders projection effects important, but then the determination of true galactocentric distances is difficult to ascertain. Thus, we estimate the galaxy's radial MG based on a simpler approach similar to \cite{Piatti+2007MNRASyoung}, D14, \cite{Parisi+2009AJclusI}, \cite{Parisi+2010AJfieldI} and C18.


We consider an elliptical system whose major axis is located along the SMC bar (i.e., along the North-East--South-West direction), with a position angle of 55.3$^{\circ}$ East of North. $a$ and $b$ are, respectively, the semi-major and semi-minor axes of the ellipse, with $a/b = 1.5$ (see D14). For each star, we estimated the value of $a$ an ellipse would have if it were centred on the SMC and aligned with the bar, and one point of its trajectory coincided with star's position. In essence, we used $a$ as a proxy for the true galactocentric distance and estimated a radial MG. We adopted annular bin widths of 0.25$^{\circ}$, and conducted linear least-squares fits. Figure \ref{fig:metgrad} shows the radial MG thus estimated for the SMC. For comparison, we also show the results from C18. For the VMC, we estimated [Fe/H] $= (-0.927 \pm 0.007) + a \times (-0.026 \pm 0.005)$ dex deg$^{-1}$ and $r=$ 0.88, out to 2.5$^{\circ}$ similar to the extent of the other two data sets. This implies a shallow but gradual metallicity distribution within the inner region, similar to the MG estimated by C18, i.e., $-0.067 \pm 0.006$ dex deg$^{-1}$ (OGLE III) and $-0.045 \pm 0.004$ dex deg$^{-1}$ (MCPS). The $Y$-axis intercepts of the MG for the three surveys are different, with the VMC appearing relatively metal-poor comapared with the MCPS and OGLE III estimates. The error bars associated with the mean metallicity in each annular radial bin suggests that the variation derived based on VMC data agrees with that of the MCPS. However, the difference is larger with respect to the OGLE III results, but it agrees within the uncertanities. Beyond 2.5$^{\circ}$, the MG gradually rises to an almost constant metal-rich ($\sim - 0.93$ dex) level from $\sim$ 3.5$^{\circ}$ to 4$^{\circ}$. The dispersion in mean metallicity values beyond 3$^{\circ}$ is relatively larger compared with the inner radii. This could be owing to small number statistics for the outer subregions and/or the metal-rich regions observed in the map.

The top panel of Figure \ref{fig:metgrad2} shows the standard deviation of [Fe/H] versus the number of subregions within each annular bin for all three data sets. We see that the values are scattered owing to small number statistics, i.e. when the number of subregions is $<$ 30. For numbers $\ge$ 30, the values are almost constant (0.05--0.06 dex). We observe some metal rich subregions ($>$ $-0.85$ dex) beyond 3$^{\circ}$ East, North-East (Wing) and very few in the West, which could be artefacts. As pointed out in Section 5.1.2, such subregions can have a scatter in the RGB's magnitude because of variations in LOS depth. Also, in such a case the CMDs cannot be completely de-contaminated from the redder part of the MW contamination. This will result in scattered colour--magnitude bins of redder colour up to the RGB tip causing shallow $|$slope$|$ estimation (hence higher metallicity). Although the number of such subregions is $\approx$ 3 per cent of the total, given the small number statistics in the outer region, they can be the reason behind the higher mean metallicity trend observed beyond 3$^{\circ}$ in the MG plot. The bottom panel shows the radial metallicity gradient estimated for all annular bins with more than 30 subregions. To get rid of such artefacts, we re-estimated the radial MG (bottom panel of Figure \ref{fig:metgrad2}) for all three data-sets by considering annular bins for $\ge$ 30 subregions. The estimated MGs are $-0.059 \pm 0.008$  and $-0.038 \pm 0.005$ dex deg$^{-1}$ out to 2$^{\circ}$ for OGLE III and MCPS respectively, whereas, for VMC we estimate a MG of $-0.031 \pm 0.005$  dex deg$^{-1}$ out to $\approx$ 2.5$^{\circ}$.  

The presence of a MG is supported by CaT metallicity estimates based on field RGB stars (\citealt{Carrera+2008AJ-CEH-SMC}, D14, \citealt{Parisi+2016AJfieldII}). \cite{Carrera+2008AJ-CEH-SMC} detected almost constant mean metallicities ($\sim -1$ dex) for their 11 (8.85$\times$8.85) arcmin$^2$ fields located within 3.0$^{\circ}$ from the centre. However, two additional fields located between 3.0$^{\circ}$ and 4$^{\circ}$ have more metal-poor values, $\sim -1.6$ dex, with relatively large error bars. The fields of \cite{Carrera+2008AJ-CEH-SMC} are distributed in the South, East and West, with none in the northern region of the SMC. These authors
suggested that the MG they derived is a result of different RG ages at different locations across the galaxy. They suggested that additional
star formation has taken place within the SMC's inner regions (within 2${^\circ}$ from the centre) in the last few billion years, leading to detection of a MG. D14 estimated a MG of $-0.075 \pm 0.011$ dex deg$^{-1}$ in the SMC's inner 5$^{\circ}$. They considered a similar geometry as we have done in this study.

The MG estimated by \cite{Parisi+2016AJfieldII}, $-0.08 \pm 0.02$ dex deg$^{-1}$ within $< 4^{\circ}$, is very similar to that of D14. The combined sample of \cite{Parisi+2010AJfieldI} and  \cite{Parisi+2016AJfieldII} (750 RGB stars distributed across 30 fields) covers small pockets within the SMC, except for the North-West region. The authors assumed an elliptical geometry for the SMC (as adopted here) but with $b/a = 0.5$.
They selected their field regions to coincide with clusters analysed by \cite{Parisi+2009AJclusI,Parisi+2015AJclusII}. Parisi et al.'s metallicity estimates of star clusters suggested a bimodality in their metallicity distribution, representing metal-poor and metal-rich components. In their figure 11, \cite{Parisi+2016AJfieldII} applied independent linear fits to both components to show that neither group exhibits a significant MG (metal-poor: $-0.03 \pm 0.05$ dex deg$^{-1}$; metal-rich: $-0.01 \pm 0.02$ dex deg$^{-1}$). Since these estimates depend on the radial extent selected, the authors issued a warning as regards the statistical significance of the two potential MGs. This was in contrast to their result using field stars. Both the clusters and their field-star studies suggest the existence of a positive MG in the outer regions ($> 4^{\circ}$). However, they also suggest that the positive MG requires a detailed follow-up study to confirm these results. 

D14 and \cite{Parisi+2016AJfieldII} attributed the MG they found to the relation between metallicity and age pertaining to the SMC. The more metal-rich stars tend to be the younger and are more centrally located within the galaxy (\citealt{DaCosta+1998AJ,Idiart+2007,Cignoni+2013ApJ}). Our estimated MG can be due to similar reasons. However, our spatially binned areas span a more homogeneous spatial distribution within the SMC compared with spectroscopic studies of individual stars, which are often confined to only small pockets within the galaxy. Thus, in estimating our mean MG, we could have smoothed out any local variation in metallicity, producing a relatively shallower gradient compared with D14 and \cite{Parisi+2016AJfieldII}.

Another factor that could have additionally contributed to the detection of a shallow MG are the ages of the population sampled. \cite{Dolphin+2001}, using WFPC2 $(HST)$ and ground-based $VI$ for a field in the outer SMC, found a metallicity of [Fe/H] = $-$1.3$\pm$0.3 dex for the oldest stars ($\ge $8 Gyr), which increased to [Fe/H] = $-$0.7$\pm$0.2 dex in the past 3 Gyr. \cite{Harris&Zaritsky2004} estimated the chemical enrichment history in the central region of the galaxy based on $UBVI$ photometry from the MCPS Survey, and found that the stars formed until $\sim$ 3 Gyr ago have a mean metallicity [Fe/H] = $-$1 dex that rises monotonically to a present value of [Fe/H] $\sim$ $-$0.4 dex. \cite{Piatti2012MNRASage-metalSMC} using Washington photometry pointed out that the SMC field stars do not show any gradients in age or metallicity. According to author, the innermost region (semi-major axis $\leq 1^{\circ}$) of the SMC hosts stellar populations that were formed $\sim$2 Gyr ago and which are more metal-rich than [Fe/H] $\sim -0.8$ dex, and are mixed along with relatively more metal-poor ([Fe/H] $\sim$ $-1.0$ to $-1.5$ dex) and older (age $\sim$ 3--8 Gyr) field stars. In our NIR metallicity map we hardly detect any region with metallicity $> -0.8$ dex. Therefore, it could also be possible that we may have selected against the younger RGB stars ($< 2$--3 Gyr-old), which are relatively metal-rich and confined to the inner SMC. Given that we observe a small colour width in our NIR CMDs, in identifying the densest RGB profile for an individual subregion we could be typically sampling the older RGB population ($> 3$ Gyr old). The metallicity variation for stellar populations younger than 3.4 Gyr from $N$-body simulations by \cite{Yozin&Bekki2014} shows a shallow gradient (their figure 15), which flattens in the outer disc ($>$3 kpc). The authors find this chemical enhancement towards the inner disc of the SMC in agreement with the Cepheid distribution (aged between 30 and 300 Myr), which is again closely correlated with recent star formation history along the SMC bar \citep{Haschke+2012aAJ}. Qualitatively, our trend agrees with these authors, although we sample a relatively older population.

Our MG is inconsistent with the derivation of \cite{Cioni2009A&Athemetallicity}, who used the C/M ratio of AGB stars as an indicator of their metallicity and found that the SMC has an almost constant [Fe/H] $\sim -1.25 \pm 0.01$ dex from its centre out to $\approx$ 11.5$^\circ$. This is rather metal-poor compared with our RGB population, and the reason for this difference requires in-depth follow-up. Note that the C/M ratios were calibrated to metallicities using RGB stars \citep[][their appendix B]{Cioni2009A&Athemetallicity}. However, we used the same indicators and calibrators. Hence, there should be no systematic uncertainties owing to age differences between our calibrators and indicators. Our MG also differs from that of \cite{Haschke+2012aAJ}, who did not find any MG using 1831 RR Lyrae stars from the OGLE III survey. Their metallicity contour map (their figure 8) shows a smooth distribution without any distinct features. Their map is restricted by the OGLE III coverage area, whereas we have obtained a map covering a much larger area. This result is also supported by \cite{Deb+2015MNRAS}. Figure 10 of \cite{Deb+2015MNRAS} shows mean metallicity values on a binned 10$\times$10 coordinate grid. Moreover, their figure 11 shows the mean metallicity distribution as a function of galactocentric distance. A shallow, tentative MG ($-0.013 \pm 0.007$ dex kpc$^{-1}$) in the de-projected plane of the SMC was detected by \cite{Kapakos&Hatz2012MNRAS} based on analysis of the $V$-band light curves of 454 RR Lyrae from OGLE III. However, the authors caution that their MG could be affected by selection effects and needs spectroscopic confirmation.


The presence of a shallow MG in the inner SMC could be related to the presence of a bar. Galactic bars are generally found in the disc regardless of the overall structure of the galaxy \citep{Zaritsky+1994}. The presence of a bar tends to flatten any MG or may create a constant metallicity distribution \citep{Martin&Roy1994}, since it induces non-circular gas motions in the disc. Several mechanisms can play a role in flattening the MGs in the outer disc, including gas inflow from tidal interactions \citep{Kewley+2010ApJ}. Episodes of close encounters occurred at 100--300 Myr and a few Gyr ago between the MCs \citep{Besla+2012, Diaz&Bekki2012}, where the first epoch corresponds to a direct collision scenario. Given that these two epochs could be recent compared with the ages of our population, the long-term effect of the bar ($>$2--3 Gyr at least) could be more significant for the evolution of a MG than the effect of a close tidal interaction. However, the SMC's structure is not that of a disc galaxy despite a possible indication of a weak bar \citep{Yozin&Bekki2014}. Simulations by \cite{Gajda+2017} follow tidally induced bars in dwarf galaxies on different orbits around a MW-like host. However, these $N$-body simulations depend on the size of the dwarf galaxy's orbit and the inclination of its disc with respect to the orbital plane. Therefore, the presence and growth of a bar in the SMC and its effects on the galaxy's MG need to be further investigated. This is beyond the scope of the present paper. 


\begin{figure}
\includegraphics[width=\columnwidth]{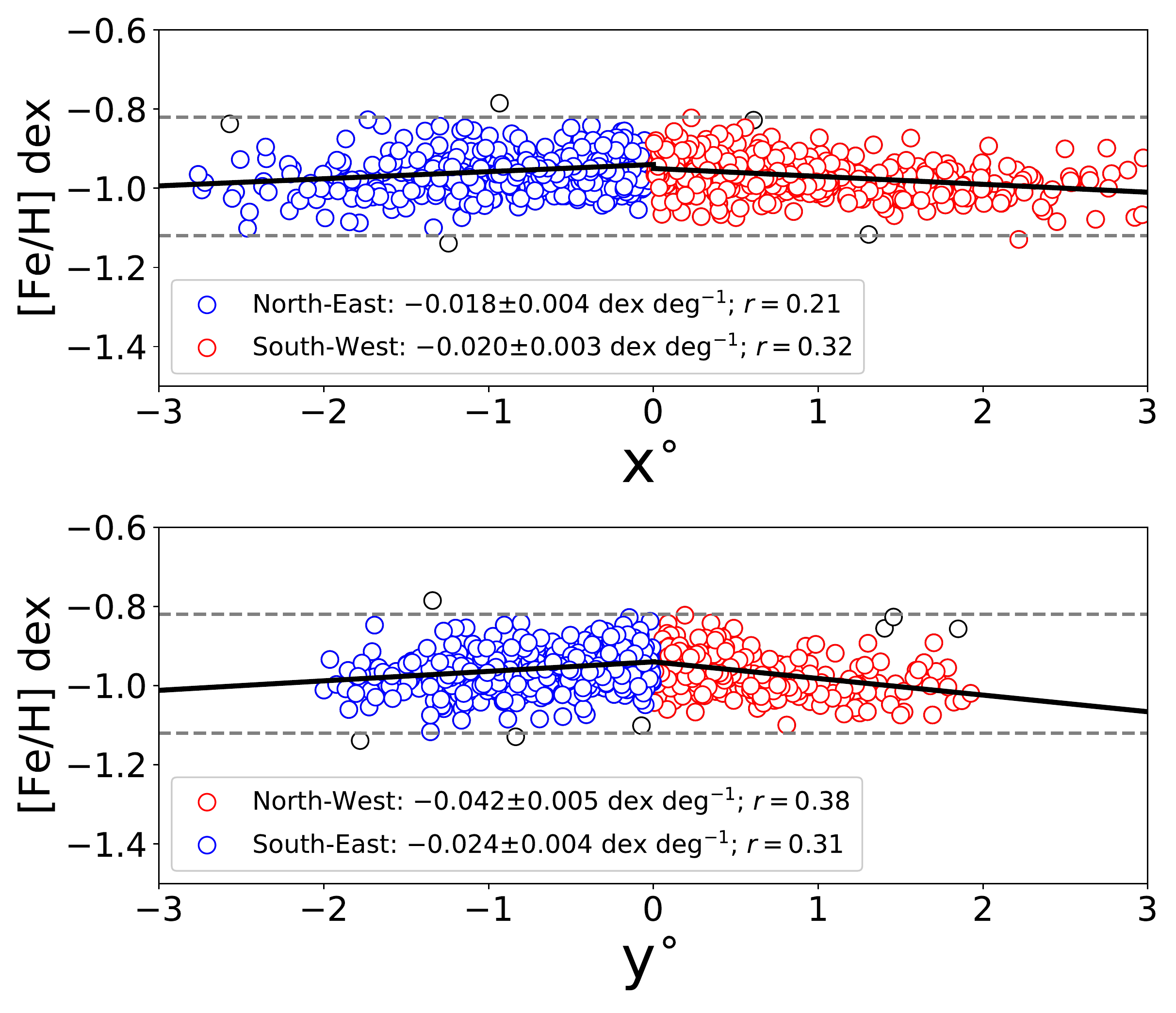}
\caption{Estimated MG along the major (top) and minor (bottom) axes. The positive major axis is along the South-West direction and the negative axis is along the North-East. Whereas, the positive minor axis is along the North-West direction and the negative axis is along the the South-East. Our data are plotted as black open circles in both panels. The red open circles denote points after estimating the MG with 3${\sigma}$ clipping along the positive major and minor axes. The blue open circles denote similar cases along the negative major and minor axes. The solid black line denotes the fitted MG for all cases. Whereas, the grey dashed lines denote the 3${\sigma}$ deviations from the mean metallicity of the SMC disc.}
\label{fig:metgrad_axes} 
\end{figure}
 
We plotted the metallicity distribution with respect to the major and minor axes (of the elliptical system) in Figure \ref{fig:metgrad_axes}, to check for any difference in the MG between the sides in the direction of the SMC (East) and that away from it (West). We have only considered points within a radius $a$ $\le$ 3$^\circ$ to avoid misinterpretation owing to any possible artefacts. The metallicity range along all axes is found to be constrained within 3${\sigma}$ of the mean [Fe/H] ($-$ 0.97 dex), with only a few deviations. The estimated MGs are: $-0.018(\pm 0.004)$ and $-0.020(\pm 0.004)$ dex deg$^{-1}$ along the North-East and South-West directions, respectively; $-0.042(\pm 0.005)$ and $-0.024(\pm 0.004)$dex deg$^{-1}$ along the North-West and South-East directions, respectively. The MGs are in general found to be shallow and similar to the radial MG. The MG along the North-East appears steeper compared with the South-West direction, but they agree within the errors. However, the MG along the North-West direction is much steeper as compared with the South-East direction. Since the eastern side of the SMC is in the direction of the LMC (and the Magellanic Bridge), our results hint at the existence of a steeper MG in directions opposite to that. This is similar to the results obtained by C18 (their figure 28). If we combine the metallicity estimates from all three surveys (OGLE III, MCPS and VMC), the radial asymmetry in MG is sustained. This could mean that the spatial metallicity distribution in the eastern part is perturbed owing to tidal interactions with the LMC. The last interaction between the MCs (100--300 Myr ago) led to the formation of the Magellanic Bridge. Studies indicate that apart from gas and young stellar populations \citep{Irwin+1990AJ,Bica+2015MNRAS} the Bridge also hosts intermediate-age/old stellar populations in its central and western part \citep{Bagheri+2013,Noel+2013ApJ,Noel+2015MNRAS,Skowron+2014}. Tidal interactions can have effect on gas as well as on the stars. \cite{Nidever+2013} and \cite{Smitha+2017MNRAS} provided evidence of a tidally stripped intermediate-age population towards the East of the SMC ($>$ 2$^{\circ}$--2.5$^{\circ}$). Even \cite{Dobbie+2014MNRAS-papI} from their extensive spectroscopy indicate substantial tidal stripping of this intermediate-age population. Thus, using a homogeneous distribution of subregions within the inner and outer SMC, our study hints at a radially asymmetric MG (within the inner 3$^\circ$), which could have resulted from its tidal interaction with the LMC.


\section{Summary}
We have successfully extended our technique of combining large-scale photometric and spectroscopic data developed by C16 and C18 for optical passbands ($V$ and $I$) to the NIR passbands ($Y$ and $K_{\rm s}$) of the VMC survey, leading to the derivation of NIR metallicity maps of the SMC. The results can be summarised as follows:
\begin{enumerate}
\item Our NIR metallicity maps exceeds the previously obtained metallicity maps of C18 in terms of area coverage (three times larger), revealing trends across 42 deg$^2$ of the SMC.
\item We estimated RGB slopes in the $Y$ versus $Y-K_{\rm s}$ CMD of $\sim$900 subregions in the SMC (within a radius of 4$^{\circ}$) and converted the slopes to metallicity values using spectroscopic data of field RGs.
\item The mean metallicity of the SMC based on VMC data is $-0.97 \pm  0.05$ dex out to a radius of 4$^{\circ}$. This agrees well with the mean metallicities estimated in previous photometric and spectroscopic studies of RGs. Some subregions at outer radii ($>$3$^{\circ}$, particularly towards the East; $\approx$ 3 per cent of the total) with metallicity $> -0.85$ dex could be artefacts. 
\item The RGB population drawn from a spatially homogeneous large-area photometric data set shows the existence of a shallow MG ($-0.031 \pm 0.005$ dex deg$^{-1}$) within 2--2.5$^{\circ}$. This is in agreement with the results from optical studies of C18. The trend at $\approx$ 3.5$^{\circ}$ to 4$^{\circ}$ hints at a flattening of the MG. We do not quantify this owing to the high dispersion in mean metallicity in the outer regions.
\item The role of a bar and/or the interaction of the LMC--SMC system in producing a shallow MG within the inner radii ($\le$ 2.5$^{\circ}$) and the flattening at the outer radii needs further investigation.
\item The MG is found to be radially asymmetric and relatively flattened towards the eastern side compared with the western side, supporting similar results by C18. We suspect that this could be due to the tidal interaction between the LMC--SMC system.
\end{enumerate}

\section*{Acknowledgements}
The authors would like to thank the Cambridge Astronomy Survey Unit (CASU) and the Wide Field Astronomy Unit (WFAU) in Edinburgh for providing the necessary data products under the support of the Science and Technology Facility Council (STFC) in the U.K. This study was based on observations made with VISTA at the La Silla Paranal Observatory under programme ID 179.B-2003. This project has received funding from the European Research Council (ERC) under the European Union’s Horizon 2020 research and innovation programme (grant agreement no. 682115). For PSF photometry, we acknowledge the use of the computer cluster at the University of Hertfordshire, Simone Zaggia for making available local computers at the University of Padova, and the ERC Consolidator Grant funding scheme (project STARKEY, grant agreement no. 615604) for supporting Stefano Rubele's work. We thank the anonymous referee for their suggestions which helped improve the clarity of the manuscript. We thank Jim Emerson and Amy Miller for their valuable comments which helped improve the manuscript.



\bibliographystyle{mnras}
\bibliography{bibliography}










\bsp	
\label{lastpage}
\end{document}